\documentclass[utf8]{frontiersSCNS} 

\usepackage{url,hyperref,lineno,microtype,subcaption}
\usepackage[onehalfspacing]{setspace}

\def\keyFont{\fontsize{8}{11}\helveticabold }
\def\firstAuthorLast{C\'orsico, A. H.} 
\def\Authors{Alejandro H. C\'orsico\,$^{1,2,*}$}
\def\Address{$^{1}$Facultad de Ciencias Astron\'omicas y Geof\'isicas, Universidad Nacional de La Plata, Paseo del Bosque s/n, (1900), La Plata, Argentina\\
$^{2}$Instituto de Astrof\'isica La Plata, IALP, CONICET-UNLP, La Plata, Argentina}
\def\corrAuthor{A. H. C\'orsico}

\def\corrEmail{acorsico@fcaglp.unlp.edu.ar}

\begin{document}
\onecolumn
\firstpage{1}

\title[White-dwarf asteroseismology with {\it Kepler}]
      {White-dwarf asteroseismology with the {\it Kepler} space telescope} 

\author[\firstAuthorLast ]{\Authors} 
\address{} 
\correspondance{} 

\extraAuth{}

\maketitle

\begin{abstract}

\section{}

In the course of their evolution, white-dwarf stars  go through at
  least one phase of variability in which the global  pulsations they undergo
allow astronomers to peer into their interiors,  this way  making possible to
  shed light on their deep inner structure and evolutionary stage by means of
  asteroseismology. The study of pulsating white dwarfs
has witnessed substantial progress in the last decade, and this has
been so largely thanks to the  arrival of continuous observations of
unprecedented quality from space,  like those of the CoRoT, {\it Kepler},
and {\it TESS} missions. This, along with the  advent of new detailed
thoretical models and the development of  improved 
asteroseismological techniques, has helped to unravel the internal
chemical structure of many pulsating white dwarfs, and, at the same
time, has opened new questions that challenge theoreticians.
In particular, uninterrupted monitoring of white-dwarf stars
for months has allowed discovering phenomena impossible to detect with
ground-based observations, despite admirable previous efforts like the
Whole Earth Telescope (WET). Here,  we start by reviewing the essential
properties of white-dwarf and pre-white dwarf stars and their pulsations,
and then, we go through the different families of pulsating objects known to
date. Finally, we review the most outstanding findings
  about pulsating white dwarfs and pre-white dwarfs
  made possible with the unprecedented-quality
  observations of the {\it Kepler} space telescope, although we envisage
  that future analyzes of space data from this mission that still await to
  be examined may reveal new secrets of these   extremely interesting
  variable stars.

\tiny
\keyFont{ \section{Keywords:} stellar evolution, white dwarf stars,
  stellar interiors, stellar oscillations, asteroseismology}
\end{abstract}

\section{Basic properties of white-dwarf and pre-white dwarf stars}

White-dwarf  stars constitute the most common compact remnants of
the evolution of stars with initial masses below $\sim 10-11
M_{\odot}$ \citep{2015ApJ...810...34W}, including our Sun.
Comprehensive accounts  referring to the origin and evolution of white dwarfs
can be found in the review articles of \cite{2001PASP..113..409F} and
\cite{2010A&ARv..18..471A}. White dwarfs are extremely old objects, with
typical ages in the range $1-10$ Gyr (1 Gyr $\equiv 10^{9}$ yr). These
stellar fossils are found in a range of masses of $0.15 \lesssim
M_{\star}/M_{\odot} \lesssim 1.25$, with an average value of
$M_{\star}/M_{\odot} \sim 0.60$ \citep{2016MNRAS.461.2100T},
although for the majority of white dwarfs the
mass lie between $M_{\star}/M_{\odot} \sim 0.50$ and
$M_{\star}/M_{\odot} \sim 0.70$. Since they are characterized by
planetary dimensions ($R_{\star} \sim 0.01 R_{\odot}$), the matter
inside is highly concentrated, with average densities of white dwarfs of
the order of $\overline{\rho}  \sim 10^6$ g/cm$^3$. Consequently,
the prevailing equation of state inside white dwarfs is that
of a highly-degenerate Fermi gas
\citep{1939isss.book.....C}.  Indeed, the hydrostatic equilibrium in
the interior of a white dwarf is provided by the pressure of degenerate
electrons counteracting the gravity force, thus avoiding the collapse
of the star. In particular, electron degeneracy
is responsible for a curious relationship between the stellar mass and radius:
the more massive the white dwarf, the smaller its radius. The process of discovery of
white dwarfs, starting from the observation of  variations of the proper motion of
Procyon and Sirius by Friedrich Wilhelm Bessel in 1844 \citep{1844MNRAS...6R.136B},
was fascinating. The emergence of the quantum-mechanics theory towards the first quarter
of the  $20^{\rm th}$ century paved the way to explain the existence of white dwarfs.
An excellent historical account  of the
discovery of white dwarfs and the theoretical arguments that led to explain
the existence of these compact objects in nature is provided by \cite{2015uswd.book.....V}.

Apart from the degenerate electrons, the interior of white dwarfs is composed
by non-degenerate ions that store the thermal energy  generated by
previous nuclear burning --- generally extinct in the white-dwarf stage.
White dwarfs
are found at a huge  interval of effective temperatures  ($4000 \lesssim
T_{\rm eff} \lesssim 200\,000$ K) thus spanning seven orders of
magnitude in luminosity, $10^{-4} \lesssim \log(L_{\star}/L_{\odot})
\lesssim 10^3$. Regarding their inner chemical composition, average-mass white dwarfs
are characterized by cores likely made of $^{12}$C and $^{16}$O ---the
products of He burning--- although ultra-massive white dwarfs
($M_{\star}/M_{\odot} \gtrsim 1.0$) could  contain cores composed of
$^{16}$O, $^{20}$Ne, and $^{24}$Mg as well, and the lowest-mass white dwarfs
(low-mass and extremely low-mass white dwarfs, abbreviated as LM and ELM white dwarfs,
respectively) could harbor cores made of $^{4}$He. In particular, ELM
white dwarfs ($M_{\star}/M_{\odot} \lesssim 0.2$) must be formed in interacting
binary systems, otherwise they would be much older than the Universe.
A description of the different formation channels of white dwarfs, either
single- or binary-star evolution, can be found in the review articles of 
\cite{2010A&ARv..18..471A} and \cite{2019A&ARv..27....7C}.  Unlike other stages
of stellar evolution, the white-dwarf phase is characterized by a relatively
simple process of gradual cooling of the star, shining at the
expense of the heat  stored in its core during the prior
evolutionary history. This process was first described in a
simplified way by \cite{1952MNRAS.112..583M}.  The cooling rate of white dwarfs
depends on many factors: surface temperature, stellar mass, core chemical composition,
surface chemical composition, etc.  The surface chemical composition
of white dwarfs defines the spectral type.  Most  white dwarfs ($\sim 80 \%$) have
atmospheres rich in $^{1}$H, and are called DA white dwarfs.  About a $\sim 15 \%$ of
white dwarfs have atmospheres rich in $^{4}$He, and are denominated DB and DO white dwarfs. The
remainder population of white dwarfs ($\sim 5 \%$) show atmospheres rich in $^{4}$He,
$^{12}$C, and $^{16}$O (PG1159 stars)\footnote{Strictly speaking, since these stars still
  have substantial nuclear burning, they should be considered pre-white dwarfs.},
$^{12}$C and $^{4}$He (DQ white dwarfs), and metals without $^{1}$H or $^{4}$He
present (DZ white dwarfs). There are also some exotic objects, such as SDSS
J0922+2928 and SDSS J1102+2054, that exhibit O in their atmospheres
\citep{2010Sci...327..188G}, SDSS J1240+6710 that has  a nearly pure $^{16}$O
atmosphere, diluted only by traces of $^{20}$Ne, $^{26}$Mg, and $^{28}$Si
\citep{2016Sci...352...67K}, and finally WD
J0551+4135, that shows a mixed $^{1}$H and $^{12}$C atmosphere \citep{2020NatAs.tmp....3H}.
The origin of these unfamiliar objects is not yet
known, but might involve stellar mergers. By virtue of the high densities and
small radii of the white dwarfs,
they are characterized by very high surface gravities, in the range
$\log g \sim 6-9$ [cm/s$^2$].   The extremely high gravities of white dwarfs
force heavy elements to sink and light nuclear species to float on the
white-dwarf surface.  This effect, called gravitational settling, explains the
purity of the atmospheres of DA and DB/DO white dwarfs. Any ``impurity'' observed
in white dwarfs atmospheres at intermediate temperatures ---for
instance, photospheric trace metals
found in many white dwarfs--- must be attributed either to accreted material
\citep[i.e., tidally disrupted planetesimals; see][]{2014A&A...566A..34K},
material floating on the surface due
to radiative levitation, or material dredged up from the interior due to convection.  The
typical chemical structure of an average-mass DA white dwarf consists of a core
 composed of $^{12}$C and $^{16}$O ---in  unknown
proportions--- containing $\sim 99 \%$ of the mass, surrounded by a
$^{4}$He layer with a mass  $M_{\rm He} \lesssim 10^{-2} M_{\star}$,
it in turn enclosed by a thin $^{1}$H envelope with a mass $M_{\rm H}
\lesssim 10^{-4} M_{\star}$.  The thickness of the $^{1}$H and
$^{4}$He mantles depends on the mass of the white dwarf. Despite the thinness
of these layers, they play a fundamental role in the cooling rate \citep{1990PhDT.........5W}. 

\begin{figure}[h!]
\begin{center}
\includegraphics[width=12cm]{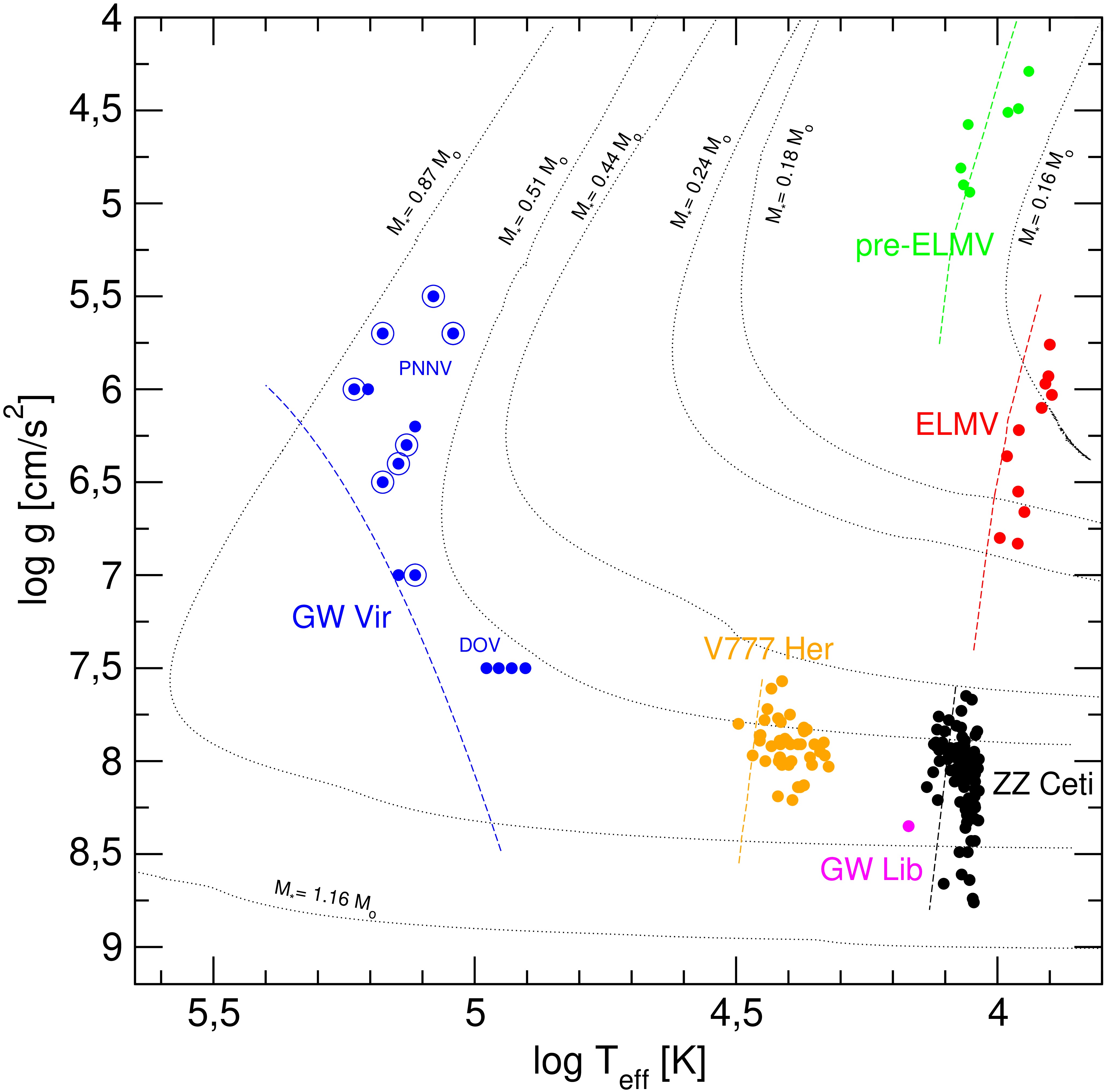}
\end{center}
\caption{The families of confirmed pulsating
  white dwarf and pre-white dwarf stars (circles of different colors) in the $\log
  T_{\rm eff}-\log g$ plane. This figure is an update of Fig. 1 of
  \cite{2019A&ARv..27....7C}. GW Vir stars indicated with blue
  circles surrounded by blue circumferences are PNNVs.  Only the
    location of the prototypical object of GW Lib class,
  the star GW Librae, has been indicated (magenta dot). Two post-VLTP (Very Late
  Thermal Pulse) evolutionary tracks for H-deficient white dwarfs \citep[$0.51$ and
  $0.87M_{\odot}$;][]{2006A&A...454..845M}, four evolutionary tracks
  of low-mass He-core H-rich white dwarfs \citep[$0.16, 0.18, 0.24$, and $0.44 M_{\odot}$;][]{2013A&A...557A..19A}, and one evolutionary track for ultra-massive
  H-rich white dwarfs \citep[$1.16 M_{\odot}$;][]{2019A&A...625A..87C} are  included
  for reference. Dashed lines correspond to the location of the theoretical hot
   boundary of the different instabilities strips.}
\label{fig:1}
\end{figure}

\begin{figure}[h!]
\begin{center}
\includegraphics[width=12cm]{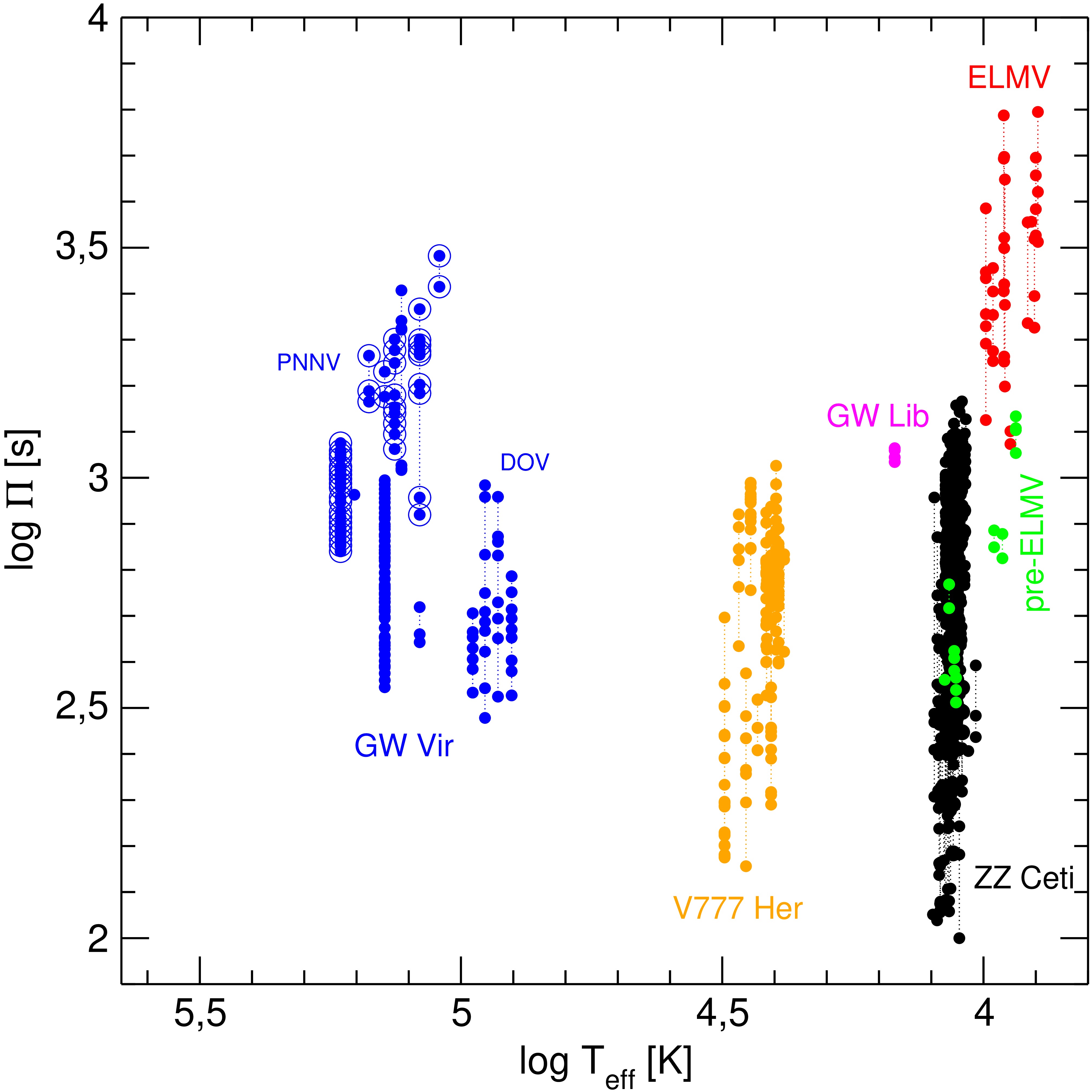}
\end{center}
\caption{The logarithm of the observed periods in terms of the logarithm
  of the effective temperature corresponding to the different families of
  confirmed pulsating white dwarf and pre-white dwarf stars (circles of
  different colors; see Fig. \ref{fig:1}). This figure is an update of
  Fig. 2 of \cite{2013EPJWC..4305005S}.}
   \label{fig:2}
\end{figure}

The population of known white dwarfs has increased dramatically in recent years.
Ground-based observations,  principally with the spectral observations of
the Sloan Digital Sky Survey \citep[SDSS][]{2000AJ....120.1579Y}, have
enlarged the number of known white dwarfs by a factor of 15
\citep{2013ApJS..204....5K,2016MNRAS.455.3413K,2019MNRAS.486.2169K}.
Recently, \cite{2019MNRAS.482.4570G} presented a catalog of $\sim
260\,000$ high-confidence white-dwarf candidates selected from {\it Gaia} DR2.

 White-dwarf research has relevant applications to various
areas of modern astrophysics.  A primary application of the analysis of
white-dwarf properties, either individual or collective ---like the
mass distribution, core chemical composition, and cooling times--- is
to place constraints on the evolution of low-mass stars, including
third dredge up and mass loss on the Asymptotic Giant Branch (AGB),
the efficiency of extra-mixing during core He burning, and nuclear
reaction rates
\citep{2002ApJ...567..643K,2003ApJ...583..878S,2003ApJ...587L..43M,2009ApJ...692.1013S,
  2016ApJ...823...46F}.

By virtue of the large number of white dwarfs that are currently
known, and thanks to the extreme longevity of these stars (they are
among the oldest objects of the Galaxy), they are a key part for
our understanding of the formation and evolution of stars,
evolution of planetary systems, and the history of our Galaxy
itself.  For instance, since the existing population of white dwarfs
holds a detailed account of the early star formation in the Galaxy,
then accurate white-dwarf luminosity functions can be  employed to deduce
the age, structure and evolution of the Galactic disk and the
nearest open and globular clusters through what is called {\it
cosmochronology}
\citep{2001PASP..113..409F,2009ApJ...697..965B,2010Natur.465..194G,2015MNRAS.448.1779B,
2013MNRAS.433..243C,
2016MNRAS.456.3729C,2016NewAR..72....1G,2017ApJ...837..162K}. 

From another perspective, since the majority of low-mass stars will
evolve into white dwarfs, then most of the host stars of planetary
systems will end their lives as white dwarfs. At present, many white
dwarfs with planetary debris  are currently being detected, and
their study provides valuable information about the chemical
composition of extra-solar planets \citep{2012MNRAS.424..333G,
2018MNRAS.477...93H}.  In particular, \cite{2019Natur.576...61G}
have detected a white dwarf accreting material probably coming from
a giant planet.

In a different context, white dwarfs are found in binary
systems. This allows to explore the interactions among stars, and in
particular, to study  type Ia supernovae progenitors
\citep{2014ARA&A..52..107M}. Indeed, since the mechanical properties
of a white dwarf are described by a Fermi gas of degenerate
electrons, there exist a limit mass ---the Chandrasekhar mass, $\sim
1.4 M_{\odot}$--- beyond which the structure of a white dwarf
becomes unstable. Therefore, a $^{12}$C/$^{16}$O-core white dwarf in
a binary system that receives mass from its companion, can
approach the Chandrasekhar mass and explode as a supernova.

Finally, another very important aspect of white dwarfs is that, due to the
high density prevailing in their interiors, they are extremely
useful as cosmic laboratories to study dense plasma physics and
solid state physics \citep[crystallization;][]
{1999ApJ...525..482M,1999ApJ...526..976M,2009ApJ...693L...6W,2019Natur.565..202T},
and ``exotic physics'' \citep[axions, neutrino magnetic dipole
moment, variation of fundamental constants,
etc;][]{1992ApJ...392L..23I,2008ApJ...682L.109I,2018MNRAS.478.2569I, 2001NewA....6..197C,
  2012MNRAS.424.2792C,2013JCAP...06..032C,2014JCAP...08..054C,
2014A&A...562A.123M,2014JCAP...10..069M}.

\section{Pulsation properties of white dwarfs and pre-white dwarfs and the classification
  of the different types of variables}
\label{pulsation_properties_classes}

An  important characteristic of white dwarfs is that, along their
evolution, all of them go through at least one phase of pulsational
instability that converts white dwarfs into variable stars. As such, variable
white dwarfs can be studied through {\it asteroseismology}, that allows to
extract key information about the internal structure of pulsating
stars through the study of their normal modes. In a sense, the global
pulsations are something like ``windows'' that allow astronomers to
``see'' inside the stars \citep{1980tsp..book.....C,1989nos..book.....U,
  2010aste.book.....A,2015pust.book.....C}.  In the case of white dwarfs, the
properties derived through asteroseismology for pulsating objects are
likely valid for non-pulsating white dwarfs as well. On the other hand, this is
the unique tool that makes it possible to delve inside the
white dwarfs. Finally, and due to the fact that the equilibrium structures of
white dwarfs are simple ---by virtue of its degenerate nature---
their pulsation properties are consequently simple and
relatively easy to model to a meaningful precision. In
particular, they can be treated within the  context of the linear theory
of stellar pulsations. These unique qualities explain why white-dwarf
asteroseismology has become in recent years one of the most powerful
tools to learn about their origin, evolution and internal structure,
complementing the traditional techniques of spectroscopy,
photometry and astrometry
\citep{2008ARA&A..46..157W,2008PASP..120.1043F,2010A&ARv..18..471A,
  2019A&ARv..27....7C}. As a matter of fact, white-dwarf asteroseismology
has been a cornerstone of the field
of asteroseismology \citep{1994ARA&A..32...37B}. Below,
we briefly describe the pulsational
properties of white dwarfs along with their immediate precursors,
the pre-white dwarfs, and the
different classes of these compact pulsators known at the time of
writing this  review (March 2020). 

Pulsations in white dwarfs manifest as brightness  fluctuations in the optical and
also in the ultraviolet (UV) and  infrared (IR) regions of the electromagnetic spectrum,
with amplitudes between 0.001 mmag and 0.4 mag. The variations are thought to be
produced by changes in the surface temperature due to
nonradial $g$ modes  \citep{1979ApJ...229..203M,1982ApJ...259..219R}
with low harmonic degree ($\ell= 1, 2$) and low and intermediate
radial order ($k$). In particular, \cite{1982ApJ...259..219R} showed
that during pulsations, the changes of the stellar radius are  fairly 
small ($\Delta R_{\star} \sim 10^{-5} R_{\star}$), and that the
variations in  surface temperature ($\Delta T_{\rm eff} \sim 200$ K)
are the true cause of the variability of pulsating white dwarfs.
\cite{1984ApJ...286..314K}
investigated the effects of $r$-mode nonradial pulsations\footnote{$r$ modes are
    toroidal modes in presence of rotation \citep{1978MNRAS.182..423P}.}
on the lightcurves and line profiles of a slowly rotating star, and
concluded that the observed variations of the prototype H-rich
pulsating white dwarf, G117$-$B15A, are not caused by $r$-mode pulsations, but
are consistent with $g$-mode pulsations. By virtue of the  high electron
degeneracy in the core of white dwarfs, the $g$-mode critical frequency --- the
``buoyancy'' or Brunt-V\"ais\"al\"a frequency--- is very low at the
core regions. This forces $g$ modes to probe particularly the regions
of the envelope \citep[see, for instance][]{1991ApJ...367..601B},
although some low radial-order modes are generally sensitive to the
central regions. In the case of hot pre-white dwarfs like PG1159 stars,
degeneracy is not as
important, and $g$ modes can propagate throughout the star, including
the core regions \citep{1985ApJ...295..547K,2006A&A...454..863C}.  The
first pulsating white dwarf ever detected, HL Tau 76, was discovered
unexpectedly by \cite{1968ApJ...153..151L}. Since then, an increasing
number of pulsating white dwarfs have been  detected, either through
observations from Earth ---mainly for candidates identified from SDSS---
and recently  as a result of uninterrupted observations of
space missions, such as
{\it Kepler}/{\it K2} \citep{2010Sci...327..977B,2014PASP..126..398H}.
Prior to the era of observations from space, an important international
collaborative tool called Whole Earth Telescope
\citep[WET;][]{1990ApJ...361..309N} was successfully used to observe
pulsating white dwarfs and pre-white dwarfs for long intervals of time without
interruption. As a matter of fact, observations from this world-wide
network of telescopes derived in the
most precise light curves obtained ever before for any pulsating star
at that time \citep[see, for
  instance,][]{1991ApJ...378..326W,1994ApJ...430..839W}.  The
properties of the light curves of pulsating white dwarfs and pre-white dwarfs are 
varied, ranging from very simple and with low-amplitude variations
(containing a single period) to  complex and with
high-amplitude fluctuations (usually harboring several periods).  The
first ones are usually associated to pulsating white dwarfs located at the hot
edge of the specific instability domain, and the last ones are found
generally for stars  populating the cool boundary of that instability
region ---as the convection zone deepens, longer-period modes are excited.
In the case of the complex lightcurves, that usually exhibit
features of nonlinearity, it is common to find harmonics and linear
combinations of eigenfrequencies that are not related to  real
pulsation modes. Rather, they are likely  connected with  physical
phenomena inherent to the outermost regions of the white dwarf, probably
related to the outer convection zone.  In the case of pulsating
pre-white dwarfs, the effective temperatures are so high that these stars
probably lack of an extensive external convective zone. This may be the
explanation for the fact that nonlinearities are not detected in any
pulsating hot pre-white dwarf star.

\begin{table}
\scriptsize
\caption{Basic characteristics of the different types of pulsating white dwarfs
  and pre-white dwarfs, sorted  by decreasing effective temperature.
  The two unconfirmed classes of pulsators are labeled with a question mark in
  parentheses in the first column.}
\begin{tabular}{llccccc}
\hline
\noalign{\smallskip}
Class    &   Year              &  $T_{\rm eff}$         & $\log g$     & Period         &   Amplitudes &  Main surface  \\
        &    of disc. (\#)     &   [$\times$ 1000 K]  & [C.G.S.]     & range [s]      &   [mag]      &  composition\\
\noalign{\smallskip}
\hline
\noalign{\smallskip}
GW Vir (PNNV) & 1984 (10)  & $100-180$  & $5.5-7$  & $420-6000$ &   $0.01-0.15$ & He, C, O \\
GW Vir (DOV)  & 1979 (9)   & $80-100$   & $7.3-7.7$  & $300-2600$ & $0.02-0.1$  & He, C, O \\
&&&&&&\\
V777 Her (DBV) & 1982 (27) & $22.4-32$  & $7.5-8.3$  &  $120-1080$   & $0.05-0.3$  & He (H)   \\
&&&&&&\\
GW Lib & 1998 (20) & $10.5-16$ & $8.35-8.7$ & $100-1900$  & $0.007-0.07$ &  H, He \\
&&&&&&\\
ZZ Cet (DAV) & 1968 (260) & $10.4-12.4$   & $7.5-9.1$  & $100-1400$           & $0.01-0.3$  & H     \\
&&&&&&\\
pre-ELMV      & 2013 (7)   & $8-13$     & $4-5$  &  $300-1400$                 & $0.001-0.05$ & He, H   \\
&&&&&&\\
ELMV          & 2012 (11)  & $7.8-10$   & $6-6.8$  & $100-6300$           & $0.002-0.044$  &  H    \\
\noalign{\smallskip}
\hline
\end{tabular}
\label{table:1}
\end{table}

Currently,  the number of known pulsating white dwarfs and
pre-white dwarfs is about 350 \citep{2019A&ARv..27....7C}.
They are distributed in six confirmed
types, namely ZZ Ceti (or DAV) stars, GW Lib stars, V777 Her (or DBV)
stars, GW Vir (or pulsating PG1159) stars, ELMV stars, and
pre-ELMV stars. There are two additional claimed classes of pulsating white dwarfs,
the hot DAV stars \citep[$T_{\rm eff} \sim 30\,000$ K, $7.3 \lesssim \log  g \lesssim
  7.8$;][]{2008MNRAS.389.1771K,2013MNRAS.432.1632K} and the DQV stars
\citep[][]{2008ApJ...678L..51M}. Pulsational instabilities in hot DAVs were predicted on
theoretical grounds by \cite{2005EAS....17..143S,2007AIPC..948...35S};
the actual existence of pulsations in these stars must be confirmed by
additional observations. On the other hand, DQVs are hot DQ (C- and He-rich atmospheres)
white dwarfs with surface parameters $19\,000\ {\rm K} \lesssim T_{\rm eff} \lesssim 22\,000$ K
and $8 \lesssim \log  g \lesssim 9$. At present, there is growing consensus
that the variability of these objects could be due to other effects than
global pulsations. The main characteristics of the
confirmed classes of pulsating white dwarfs are
listed in Table \ref{table:1} and the location in the $\log T_{\rm
  eff}-\log g$ is depicted in Fig. \ref{fig:1} \citep[details can be
  found in][]{2019A&ARv..27....7C}. In Table \ref{table:1}, the second
column indicates the discovery year of the first  star of each type
and the number of detected objects at the time of writing this review,
the third column corresponds the range of effective temperatures at
which they are found (instability domain), the fourth column
shows the range of surface gravity, the fifth column indicates the
range of periods detected, the sixth column is the range of
amplitudes of the fluctuations in the light curves, and the seventh
column shows the surface chemical composition.  Pulsation periods usually span
the interval $\sim 100-1400$ s; however PNNVs and ELMVs display longer
periods, up to $\sim 6300$ s. In Fig. \ref{fig:2} we show the periods
detected in the different families of pulsating white dwarfs and
pre-white dwarfs. Curiously, and due to the compactness of these objects,
the periods of $g$ modes in white dwarfs  have magnitudes comparable to
the values of the periods of $p$ modes exhibited by non-degenerate pulsating stars,
like the Sun.

Pulsations in white dwarfs are self-excited, in contrast to the stochastic
pulsations, that are forced oscillations driven by turbulent convection in the
Sun, the solar-like, and red giant pulsating stars. The cause of the pulsations in white dwarfs
appears to be a mechanism that begins to act  when the star cools down to an
effective temperature at which the dominant nuclear species becomes
partially ionized near the stellar surface (that is, when there is a
coexistence of neutral and ionized atoms of the same chemical element;
``partial ionization zone'').  At the partial ionization zones, the
opacity increases at the compression phase of the oscillations, and this
excites global pulsations with periods that are of the same order of magnitude
that the thermal timescale at the driving region.
The $\kappa-\gamma$ mechanism has been invoked to
explain at least the onset of pulsations in all the categories of
pulsating white dwarfs \citep[see, for instance,][for the case of the DAV and
  GW Vir stars,
  respectively]{1982ApJ...252L..65W,1984ApJ...281..800S}. For
DAVs and DBVs, the validity of the $\kappa-\gamma$ mechanism has
been questioned. Instead, it has been proposed the so-called
``convective driving'' mechanism
\citep{1991MNRAS.251..673B,1999ApJ...511..904G}, that becomes efficient
  when the base of the outer convective zone sinks, as the responsible of mode
driving.  For GW Vir stars,  which do not have surface convection zones as
  a result of their very-high effective temperatures, the
  $\kappa-\gamma$ mechanism seems to be enough to destabilize high-order $g$-mode
  pulsations\footnote{There exist theoretical evidence of another mechat could
  excite global pulsations in white dwarfs and pre-white dwarfs, that is, 
  the $\epsilon$ mechanism due to nuclear burning. This could be responsible
  for short-period $g$-mode pulsations in GW Vir stars \citep{2009ApJ...701.1008C},
  ELMVs \citep{2014ApJ...793L..17C},
  ZZ Ceti stars coming from low-metalicity progenitors \citep{2016A&A...595A..45C},
  and in very hot DA white dwarfs \citep{2014PASJ...66...76M,2017EPJWC.15206012C}.
  However, the predicted pulsations have not been detected so far in any case.}.
The proposed mechanisms are able to predict
with decent precision  the  $T_{\rm eff}$ of the blue
(hot) edge of the instability strips. However, in general all
theories fail in terms of the predicted effective temperature of the
red (cool) boundary, where white dwarfs quench their pulsations. This is
discouraging, since even detailed calculations that take into account
the interaction between convection and pulsations \citep{2012A&A...539A..87V} fail to
reproduce the location of the red edges in the cases of DAVs and
DBVs. Recently, \cite{2018ApJ...863...82L} have proposed an
theoretical framework that accounts for the existence of the red egde
of the DAV instability strip.

Therefore, the physical processes that excite and quench  pulsations
in white dwarfs are not fully  understood. This happens in general for all
the classes of pulsating stars, aside from white dwarfs. However, this does
not prevent us to extract information from the pulsation periods
through adiabatic asteroseismological analyses alone. In
the case of white dwarfs,  these analyses are based on the forward
approximation, in which the  measured periods are compared with the
periods calculated on a  huge set of white-dwarf models to obtain a representative
asteroseismological solution of the star under study.  The theoretical
models can be the result of detailed computations that take into
account the complete evolutionary history of the white-dwarf progenitor, the approach
adopted by the La Plata Group\footnote{{\tt http://evolgroup.fcaglp.unlp.edu.ar/}}
\citep[see, e.g.,][]{2008A&A...478..869C,2012A&A...541A..42C}, or instead,
they can consist of static white-dwarf models with parameterized chemical composition
profiles \citep[see, for instance][]{2018Natur.554...73G}. A variant
of this last method is the use of evolutionary models with
parameterized chemical profiles
\citep[see][]{2019ApJ...871...13B}. These approaches are complementary
among them \citep{2019A&ARv..27....7C}. In some cases, a large set
of consecutive periods can be detected, and it is possible to make
asteroseismological inferences of the stellar mass from the mean
period spacing. This is the case of several DBV and GW Vir stars.

\section{Pulsating white dwarfs monitored by the {\it Kepler} spacecraft}

\begin{table}
  \caption{All the published pulsating white-dwarf stars observed with {\it Kepler}
    spacecraft ({\it Kepler} and {\it K2} missions). The values of $T_{\rm eff}$,
    $\log g$, ans $M_{\star}$ are those published in \cite{2019A&ARv..27....7C}, except
    in the case of EPIC 211891315, EPIC 220329764, EPIC 220453225, and EPIC 228952212,
    for which the values are those of \cite{2017PhDT........14C}. The
    last column corresponds to the paper in which the pulsations were analyzed.}  
\begin{center}
\begin{tabular}{lccccccl}
\hline
\hline
\noalign{\smallskip}
Star                    & Class      & $T_{\rm eff}$ [K]  & $\log g$ & $M_{\star}$ [$M_{\odot}$] & $P_{\rm rot}$ [h] &
Mission  & Reference  \\
\hline
\noalign{\smallskip}
KIC 4357037             & DAV &   $12\,650$ & $8.01$ &  $0.62$ & 22.0 & {\it Kepler}  & \cite{2017ApJS..232...23H} \\
KIC 4552982$^{\ddag}$    & DAV &   $10\,860$ & $8.16$ &  $0.71$ & 18.4 & {\it Kepler}  & \cite{2015ApJ...809...14B} \\
KIC 7594781             & DAV &   $11\,730$ & $8.11$ &  $0.67$ & 26.8 & {\it Kepler}  & \cite{2017ApJS..232...23H} \\ 
KIC 10132702            & DAV &   $11\,940$ & $8.12$ &  $0.68$ & 11.2 & {\it Kepler}  & \cite{2017ApJS..232...23H} \\
KIC 11911480            & DAV &   $11\,580$ & $7.96$ &  $0.58$ & 74.7 & {\it Kepler}  & \cite{2014MNRAS.438.3086G} \\  
EPIC 60017836           & DAV &   $10\,980$ & $8.00$ &  $0.57$ &  6.9 & {\it K2}      & \cite{2014ApJ...789...85H} \\   
EPIC 201355934          & DAV &   $11\,770$ & $7.97$ &  $0.59$ & $\hdots$ & {\it K2}      & \cite{2017ApJS..232...23H} \\
EPIC 201719578          & DAV &   $11\,070$ & $7.94$ &  $0.57$ & 26.8 & {\it K2}      & \cite{2017ApJS..232...23H} \\ 
EPIC 201730811          & DAV &   $12\,480$ & $7.96$ &  $0.68$ &  2.6 & {\it K2}      & \cite{2015MNRAS.451.1701H} \\
EPIC 201802933          & DAV &   $12\,330$ & $8.11$ &  $0.68$ & 31.3 & {\it K2}      & \cite{2017ApJS..232...23H} \\
EPIC 201806008$^{\ddag}$  & DAV &   $10\,910$ & $8.02$ &  $0.61$ & 31.3 & {\it K2}      & \cite{2017ApJS..232...23H} \\
EPIC 206212611          & DAV &   $10\,830$ & $8.00$ &  $0.60$ & $\hdots$ & {\it K2}      & \cite{2017ApJS..232...23H} \\
EPIC 210377280          & DAV &   $11\,590$ & $7.94$ &  $0.57$ & $\hdots$ & {\it K2}      & \cite{2017ApJ...851...24B} \\ 
EPIC 210397465          & DAV &   $11\,200$ & $7.71$ &  $0.45$ &  49.1 & {\it K2}      & \cite{2017ApJS..232...23H} \\
EPIC 211596649          & DAV &   $11\,600$ & $7.91$ &  $0.56$ &  81.8 & {\it K2}      & \cite{2017ApJS..232...23H} \\
EPIC 211629697$^{\ddag}$ & DAV &   $10\,600$ & $7.77$ &  $0.48$ &  64.0 & {\it K2}      & \cite{2016ApJ...829...82B} \\ 
EPIC 211891315$^{\S}$    & DAV &  $11\,310$ & $8.03$ &  $\hdots$   & $\hdots$ & {\it K2}      & \cite{2017PhDT........14C} \\ 
EPIC 211914185          & DAV &   $13\,590$ & $8.43$ &  $0.88$ &  1.1 & {\it K2}      & \cite{2017ApJS..232...23H} \\
EPIC 211916160          & DAV &   $11\,510$ & $7.96$ &  $0.58$ & $\hdots$ & {\it K2}      & \cite{2017ApJS..232...23H} \\ 
EPIC 211926430          & DAV &   $11\,420$ & $7.98$ &  $0.59$ &  25.4 & {\it K2}      & \cite{2017ApJS..232...23H} \\
EPIC 220204626          & DAV &   $11\,620$ & $8.17$ &  $0.71$ &  24.3 & {\it K2}      & \cite{2017ApJS..232...23H} \\
EPIC 220258806          & DAV &   $12\,800$ & $8.09$ &  $0.66$ &  30.0 & {\it K2}      & \cite{2017ApJS..232...23H} \\
EPIC 220274129          & DAV &   $11\,810$ & $8.03$ &  $0.62$ & 12.7 & {\it K2}      & \cite{2017ApJ...851...24B} \\  
EPIC 220329764$^{\ddag}$  & DAV &   $11\,180$ & $8.03$ &  $0.62$ &$\hdots$ & {\it K2}      & \cite{2017PhDT........14C} \\ 
EPIC 220347759          & DAV &   $12\,770$ & $8.08$ &  $0.66$ &  31.7 & {\it K2}      & \cite{2017ApJS..232...23H} \\
EPIC 220453225$^{\ddag}$  & DAV &   $11\,220$ & $8.04$ &  $0.62$ & $\hdots$ & {\it K2}      & \cite{2017ApJS..232...23H} \\ 
EPIC 228682478     & DAV &   $12\,070$ & $8.18$ &  $0.72$ & 109.1 & {\it K2}      & \cite{2017ApJS..232...23H} \\
EPIC 228952212$^{\ddag}$  & DAV &   $11\,080$ & $7.95$ &  $0.58$ &$\hdots$ & {\it K2}      & \cite{2017PhDT........14C} \\ 
EPIC 229227292$^{\ddag}$  & DAV &   $11\,210$ & $8.03$ &  $0.62$ &  29.4 & {\it K2}      & \cite{2016ApJ...829...82B} \\
EPIC 229228364$^{\ddag}$  & DAV &   $11\,030$ & $8.03$ &  $0.62$ & $\hdots$ & {\it K2}      & \cite{2017ApJS..232...23H} \\ 
EPIC 229228478          & DAV &   $12\,500$ & $7.93$ &  $0.57$ & $\hdots$ & {\it K2}      & \cite{2017ApJS..232...23H} \\
EPIC 229228480          & DAV &   $12\,450$ & $8.18$ &  $0.72$ & $\hdots$ & {\it K2}      & \cite{2017ApJS..232...23H} \\
\noalign{\smallskip}
\hline
\noalign{\smallskip}
KIC 8626021        & DBV &   $29\,700$ & $7.890$ & $0.56$ & 43.0  & {\it Kepler}   & \cite{2011ApJ...736L..39O} \\
PG 0112+104        & DBV &   $31\,040$ & $7.800$ & $0.58$ & 10.2 & {\it K2}       & \cite{2017ApJ...835..277H} \\
\noalign{\smallskip}
\hline
\hline
\end{tabular}
\end{center}
\begin{center}
$^{\ddag}$ Outbursting ZZ Ceti stars. $^{\S}$ Probable outbursting ZZ Ceti star.
\end{center}
\label{table:2}
\end{table}

The primary goal of the {\it Kepler} spacecraft was to detect
transiting  planets, discovering by  the end of the mission a total of
more than $4\,000$ exoplanets \citep{2018ApJS..235...38T}.
As a byproduct of the planet hunt,
high-quality photometric data of variable stars was collected,
allowing a rapid blooming of asteroseismology of many classes of
pulsating stars, in particular solar-like and red giant pulsators.
The initial survey of pulsating stars undertaken by
the {\it Kepler} Asteroseismic Science Consortium (KASC) contained a total
of 113 compact pulsator candidates (including subdwarfs and white dwarfs) which
were checked for variability using {\it Kepler} short-cadence
exposures. Unfortunately, none of the 17 white dwarfs contained in the sample
turned out to be variable.  In order not to miss the exceptional
capacity of the {\it Kepler} mission, other candidate objects
\citep[not included in the initial survey;][]{2011MNRAS.414.2860O}
began to be  monitored with {\it Kepler}.
Fortunately, this search resulted in the first observations of pulsating
white dwarfs by this space mission \citep{2011ApJ...736L..39O,2011ApJ...741L..16H,2014MNRAS.438.3086G,
  2015ASPC..493..169G,2016MNRAS.457.2855G}. Once it finished the nominal 4-year mission,
the {\it Kepler} spacecraft lost two of its four reaction wheels
(2013 May). The mission was then redirected to monitoring new fields along
the ecliptic plane during uninterrupted time-intervals of
$\sim 90$ days \citep{2014PASP..126..398H}.
This extended  {\it Kepler} mission was renamed as {\it K2}
and operated from 2014 until the spacecraft ran out of fuel in 2018.
The {\it K2} mission provided an exceptional chance to extend space-based
monitoring  for a lot of new white dwarfs. This extended mission
was cleverly exploited by astronomers, and,  at the end, many
pulsating white dwarfs were discovered and studied.

In the following sections, we summarize the main findings achieved in white-dwarf
asteroseismology thanks to the observations of {\it Kepler} and {\it K2}.  In
particular, we focus on several detailed asteroseismological studies
performed on pulsating white dwarfs observed by this space mission, and also on
selected outstanding discoveries that have been the result of
uninterrupted observations from space.  As a result of the {\it Kepler}
and {\it K2} observations, 2 DBV stars and 29 DAV stars have been intensively
studied, and the results of those analyses have
been published. We have to add to this list
3 more DAV stars analyzed and published in a PhD Thesis \citep{2017PhDT........14C}.
One of the DBV stars, KIC 8626021,
has been the focus of intense modeling by several independent groups of
asteroseismologists, thanks to which it has been possible to know its
internal structure with unprecedented precision.  In Table \ref{table:2} we show the
main characteristics of all the published pulsating white dwarfs observed with the
{\it Kepler} spacecraft (both the {\it Kepler} and {\it K2} missions). The first
column corresponds to the name of the star, the second one is the class of pulsating white dwarf,
the third and fourth columns correspond to the effective temperature and surface gravity
values, the fifth column is the stellar mass, the sixth correspond to the
rotation period determined using the observed splittings of frequencies, the seventh
column gives the phase of the space mission ({\it Kepler} or {\it K2}), and the
last column shows the reference to the paper in which the pulsational analysis
has been presented.

 The findings reported from here to the end of the article
  are associated to results already published (up to early
  2020). However, we emphasize that
  there is still a large amount of reduced data from the {\it Kepler} and {\it K2}
  missions that have not yet been analyzed, and that could translate into future
  new discoveries about pulsating white-dwarf and pre-white-dwarf stars.

\subsection{A V777 Her star in the nominal {\it Kepler} mission field: KIC 8626021}
\label{KIC8626021}

\begin{figure}[h!]
\begin{center}
\includegraphics[width=14cm]{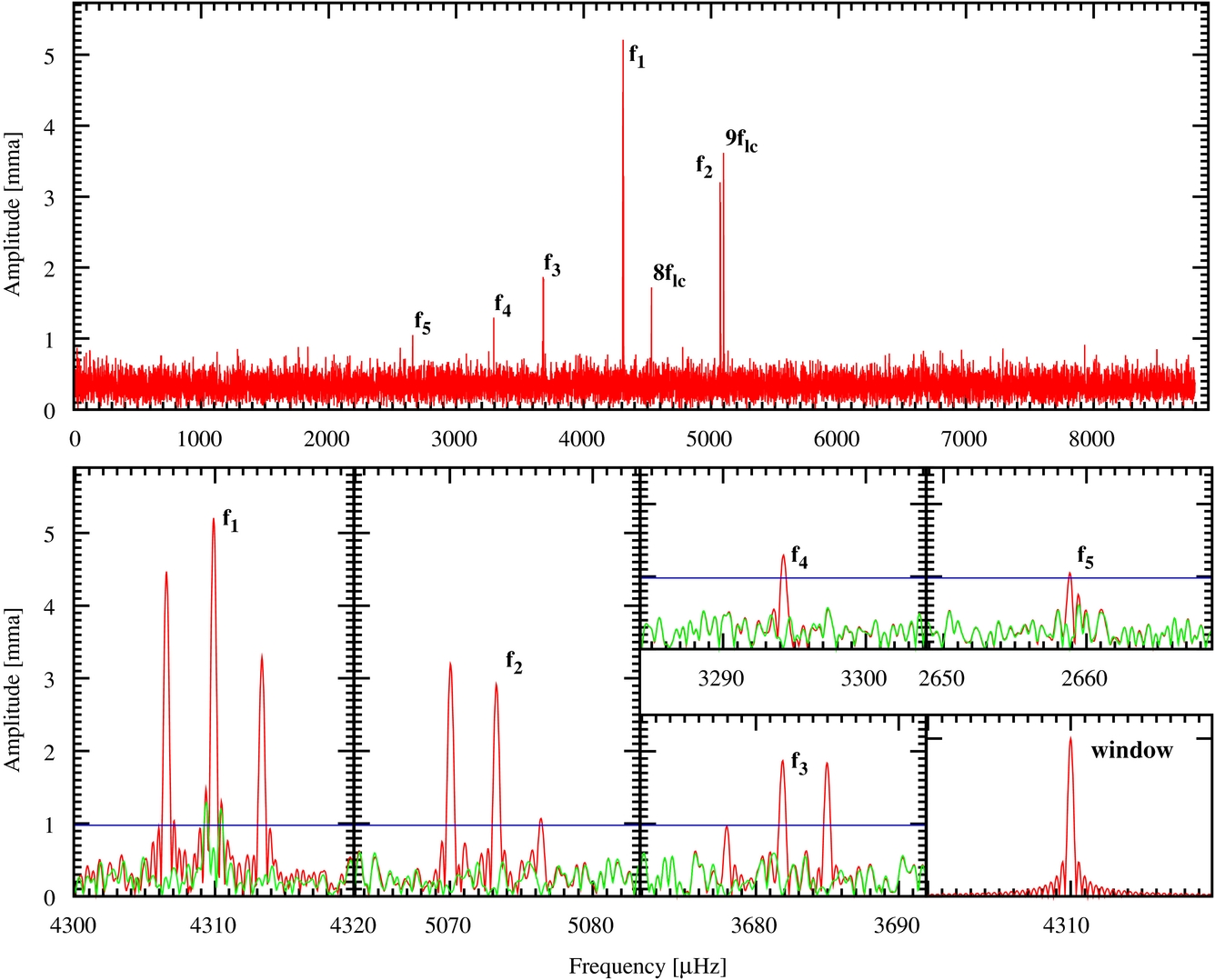}
\end{center}
\caption{Fourier Transform of the {\it Kepler} lightcurve of
  the DBV star KIC 8626021 (upper panel) derived by \citep{2011ApJ...736L..39O}.
  The peaks labeled as $8f_{\rm lc}$ and $9f_{\rm lc}$
  are two long-cadence artifacts. The real frequencies are labeled as $f_1$ to
  $f_5$. Lower panels: Zoom-in views of the five frequencies, including
  the spectral window of the {\it Kepler} data set (final panel). Reproduced by
  permission of the AAS.}
   \label{fig:3}
\end{figure}

The DB white dwarf star GALEX J192904.6+444708 ($T_{\rm eff}= 24\,900\pm750$ K
and $\log g=7.91\pm0.07$), identified  with the name
KIC 8626021 in the {\it Kepler} input catalog (also known as WD J1929+4447),
was photometrically observed by \cite{2011ApJ...736L..39O} to check
for variability from the Earth. This star was part of a number of
targets that had not been included in the original sample of 17 white dwarfs to
be observed by {\it Kepler}. A 2-hr photometry run on this star clearly
revealed the presence of variability with a period of 232 s
and an additional period around 270 s. With ground-based photometry,
the dataset  was  too  short  to  perform  any  meaningful
analysis, but it clearly demonstrated the pulsational nature of
KIC 8626021. After that, \cite{2011ApJ...736L..39O}
analyzed 1 month of short-cadence observations of this pulsating
star by {\it Kepler}. The Fourier transform of the lightcurve is shown
in the upper panel of Fig. \ref{fig:3}. The detected frequencies are
depicted in the sub-panels, three of which showing clear signals of   
triplet structures due to rotation. The periods detected appear to be
associated to a sequence of $\ell= 1$ $g$ modes with low radial order
values. The period spacing is close to 36 sec, typical of DBV
stars. On the basis of the frequency splitting
($\delta \nu \sim 3.3$ $\mu$Hz) it is possible to infer
an estimate of the rotational period of $P_{\rm rot} \sim 1.7$ days. 
By using the grids of model atmospheres of \cite{2010MmSAI..81..921K},
\cite{2011ApJ...736L..39O} determined $T_{\rm eff}= 24\,900\pm750$ K
and $\log g=7.91\pm0.07$, placing the star in the middle of the DBV instability
strip. 

Shortly after the discovery of pulsations in KIC 8626021,
the first detailed asteroseismological analysis of the star was performed by
\cite{2011ApJ...742L..16B}. They presented an analysis
based on the five-mode pulsation spectrum detected by \citep{2011ApJ...736L..39O}
with {\it Kepler} data.  The observed pulsational characteristics of the star
and the asteroseismic analysis strongly suggested that KIC 8626021
was actually hotter than indicated by model-atmosphere fits to the low
signal-to-noise electromagnetic spectrum of the object. Indeed, \cite{2011ApJ...742L..16B} employed
a large set of evolutionary models of DB white dwarfs with parametrized chemical profiles,
and took three different avenues to determine
the effective temperature of KIC 8626021: (i) by means of an  inspection
of the observed pulsation spectrum, noting that only short-period $g$ modes were present;
(ii) employing  the average separation between consecutive periods; and (iii) by
carrying out asteroseismological
perio-to-period fits of the pulsation spectrum. All three approaches pointed to an effective
temperature of $T_{\rm eff}\sim 29\,200$ K, in disagreement with the spectroscopic
value derived by \cite{2011ApJ...736L..39O} ($T_{\rm eff}= 24\,900$ K).
If true, since hot DBVs are  thought to lose a large portion of their internal energy
through the emission of plasmon neutrinos \citep{2004ApJ...602L.109W},
this star could be a extremely important target to place limits on the
plasmon-neutrino emission rate. However, the measurement of a rate of
period change requires of extremely stable (over many years) oscillation
modes, a condition that is not perfectly met by
KIC 8626021 (see at the end of this Sect.).

\begin{figure}[h!]
\begin{center}
\includegraphics[width=12cm]{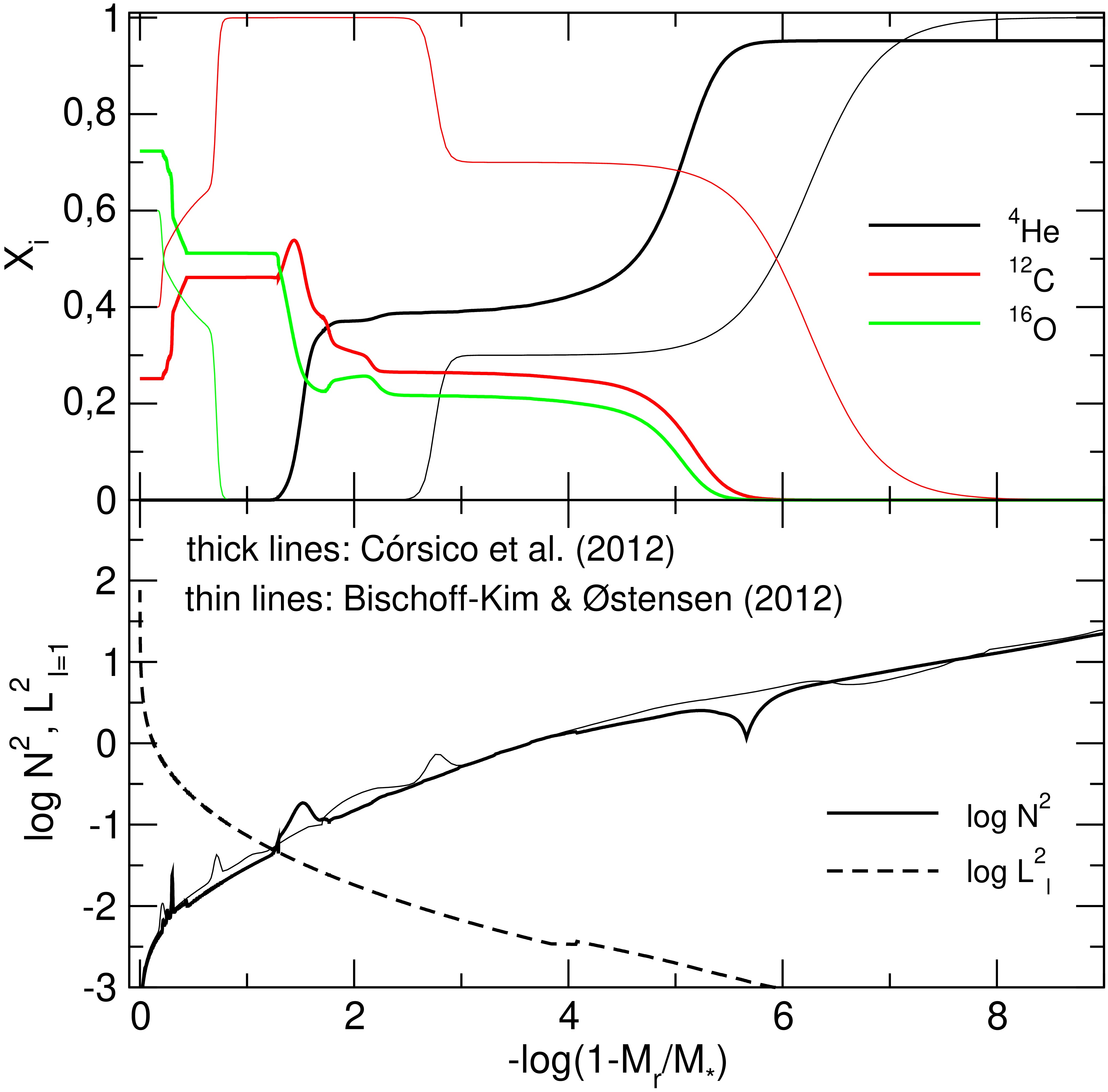}
\end{center}
\caption{Internal  chemical profiles (upper panel), and the logarithm of
  the squared Brunt-V\"ais\"al\"a and Lamb frequencies for $\ell= 1$
  (lower panel)  corresponding  to  a template  DB white-dwarf model of
  \cite{2012A&A...541A..42C} with a  stellar  mass $M_{\star}= 0.565M_{\odot}$,
  an  effective  temperature $T_{\rm eff} \sim 28\,400$ K, and a
  He envelope mass of $M_{\rm He}/M_{\star} \sim 6.7 \times 10^{-3}$ (thick lines).
  For comparison, we show with thin lines the chemical profiles and
  the Brunt-V\"ais\"al\"a frequency of the best-fit model for KIC 8626021
  found by \cite{2011ApJ...742L..16B}.}
   \label{fig:4}
\end{figure}

As a sanity check for this important result, an independent asteroseismological
analysis was crucial at that time. This was not long in coming.
\cite{2012A&A...541A..42C} carried out a second asteroseismological analysis,
which was based on a set of fully evolutionary/pulsational DB white-dwarf models
constructed with the {\tt LPCODE} evolutionary code \citep{2005A&A...440L...1A} and
the {\tt LP-PUL} pulsation code \citep{2006A&A...454..863C}.
By considering the  mean period  spacing  of KIC 8626021,  \cite{2012A&A...541A..42C} found  that
the  star should be substantially more massive than suggested by spectroscopy.
From period-to-period fits these authors found an asteroseismological
model characterized by an effective temperature much higher than the
spectroscopic estimate, in agreement with the results of
\cite{2011ApJ...742L..16B}.  So, this analysis was the second piece of evidence
that  KIC 8626021 should be located near the blue edge of the DBV
instability strip. A very interesting and exciting point is that the
DB white-dwarf models used by both groups were completely different, particularly
regarding the composition profiles (see Fig. \ref{fig:4}),
but nonetheless, similar conclusions were reached regarding the effective
temperature of the star.

A subsequent asteroseismological analysis of KIC 8626021 was carried
out by \cite{2014ApJ...794...39B} on the basis of an augmented set of
observed pulsation periods and DB white-dwarf evolutionary models with
parametrized chemical profiles, like in
\cite{2011ApJ...742L..16B}. The analysis of \cite{2014ApJ...794...39B}
was better constrained by the employment of a set of 7 independent
periods for this star ---instead of the 5 original periods---
resulting from 2-yr observations of {\it Kepler}.  By exploiting the
presence of the various triplets and doublets, the authors  were able
to constrain the value of $\ell$ and $m$, something that simplified
the period-to-period fit and reduced the assumptions at the
outset. The resulting asteroseismological model for  KIC 8626021 is
characterized by a thin pure-He envelope ($\log(1-M_{\rm
  He}/M_{\star}= -7.90$) and a high effective temperature ($T_{\rm
  eff}= 29\,650$ K). This last result confirmed that the star is near
the blue edge of the DBV instability strip, in line with the previous
analyses.  Also, a thin pure-helium layer constituted evidence that
the hotter the DB model, the thinner the pure-He layer, supporting the
theory that as a DB cools and He diffuses outward, the pure-He layer
becomes thicker \cite[see][and references
  therein]{2014ApJ...794...39B}.

The most detailed asteroseismological analysis on  KIC
8626021 was presented by \cite{2018Natur.554...73G}. These authors
considered a set of 8 periods from the analysis of
\cite{2016A&A...585A..22Z}. The additional period comes from the
alternative interpretation that the structure in the $3.677-3.686\
\mu$Hz range of the spectrum, identified previously as a triplet of
frequencies, is actually a doublet, and then the third component is
actually an independent mode. \cite{2018Natur.554...73G} presented  a
new method for parameterizing the chemical profiles in the core
of static white-dwarf models, based on Akima splines.  They  derived an
asteroseismological model characterized by a $^{16}$O content and an
extent of its core that clearly go beyond the limits predicted by
standard DB white dwarf models resulting from fully evolutionary computations.
The asteroseismological
model is characterized by $T_{\rm eff}= 29\,968\pm200$ K and $\log
g=7.917\pm0.009$, which closely match the independent measurements
obtained from spectroscopy by the same authors ($T_{\rm eff}=
29\,360\pm780$ K and $\log g= 7.89\pm0.05$) using the DB white-dwarf model
atmospheres of \cite{2011ApJ...737...28B}.  One point to note is that
these results confirm the predictions of previous studies that this
star is a hot DBV and is close to the blue edge of the V777 Her
instability strip.  On the other hand, with the new spectroscopic
determination of $T_{\rm eff}$, the discrepancy between
asteroseismology and spectroscopy is eliminated.  But the most
relevant conclusion of this work is that the star seems to have a
$^{16}$O core (and thus a $^{16}$O content) much larger than what 
standard evolutionary calculations predict. The total $^{16}$O content
of the white-dwarf core reaches $78.0\pm4.2\%$,  large in excess as compared
  with the expected value of around $64 \%$ for a standard evolutionary DB white-dwarf model of the
same stellar mass. This result constitutes a challenge for the theory of white-dwarf
formation and has aroused the interest of several research groups to
try to explain how an object of these characteristics could be
formed. For example, \cite{2019A&A...630A.100D}  find that,  within our
current understanding of white-dwarf formation and evolution, it is hard
(if not impossible) to replicate the most relevant features of the chemical
structure of KIC 08626021 derived with asteroseismology.
\cite{2018ApJ...867L..30T} have drawn the attention
  about the impact that 
neutrino emission ---a physical ingredient ignored in the DB white-dwarf modeling on which
the \cite{2018Natur.554...73G} analysis is based--- should have on the pulsation periods of KIC 08626021.  In order to
assess the effect of neutrino emission on the asteroseismological
solution of \cite{2018Natur.554...73G},
\cite{2019A&A...628L...2C} have redone the analysis on this star
by incorporating the effects of neutrino
cooling, and basically find the same asteroseismological model for
KIC 08626021 than in \cite{2018Natur.554...73G}.  A summary of all the
asteroseismological analyses carried out until now for KIC 08626021 are
given in Table \ref{table:3}. 

\begin{table}
\centering
\caption{The main  characteristics of  KIC 08626021. The  second column
  corresponds to the spectroscopic results by {\O}EA11 \citep{2011ApJ...736L..39O},
  the third, fourth, fifth, sixth, and seventh columns present the results  from
  the  asteroseismological studies of  BK{\O}11 \citep{2011ApJ...742L..16B},
  CEA12 \citep{2012A&A...541A..42C},
  BKEA14 \citep{2014ApJ...794...39B},
  GEA18 \citep{2018Natur.554...73G},
  and CHEA19  \cite{2019A&A...628L...2C}, respectively. The last two rows correspond to
the coordinate of the basis of the pure-He envelope, and the basis of the mixed zone containing
$^4$He, $^{12}$C and $^{16}$O (see Fig. \ref{fig:4}).}
\begin{tabular}{lcccccc}
\hline
\hline
\noalign{\smallskip}
Quantity                    & {\O}EA11             & BK{\O}11     & CEA12  & BKEA14 & GEA18  & CHEA19\\
\hline
\noalign{\smallskip}
$T_{\rm eff}$ [K]              & $24\,950 \pm 750$  &  $29\,200$     & $27\,263$ & $29\,650$  & $29\,968$ & $30\,114$  \\
$M_{\star}$ [$M_{\odot}$]       & $0.56\pm 0.03$     &  $0.570$       & $0.664$   & $0.550$    & $0.570$   & $0.562$    \\ 
$\log g$ [cm/s$^2$]          & $7.91 \pm 0.07$    &  ---           & $8.099$   & ---        & $7.917$   & $7.905$    \\ 
$\log (L_{\star}/L_{\odot})$    & ---                &  ---           & $-1.14$   & ---        &  $-0.86$  &   ---      \\  
$\log(R_{\star}/R_{\odot})$     & ---                &  ---           & $-1.93$   & ---        & $-1.86$   &  $-1.86$   \\  
\noalign{\smallskip}
\hline
\noalign{\smallskip}
$X_{\rm O}$ (center)          & ---                & $0.60-0.65$    & $0.65$    & $0.55$     & $0.86$    &  $0.84$  \\
$\log(1-M_{\rm He}/M_{\star})$  & ---                & $-6.30$        & $-5.95$   & $-7.90$    & $-7.63$   &  $-7.83$ \\ 
$\log(1-M_{\rm env}/M_{\star})$ & ---                & $-2.80$        & $-1.63$   & $-3.10$    & $-3.23$   &  $-3.48$ \\
\noalign{\smallskip}
\hline
\hline
\end{tabular}
\label{table:3}
\end{table}

We close this section by noting an interesting phenomenon discovered
thanks to the continuous observations of KIC
08626021 with {\it Kepler}. \cite{2016A&A...585A..22Z}  have detected amplitude and
frequency modulations of the components of the triplets of frequencies
due to rotation present in the pulsation spectrum of KIC
08626021. Similar secular
changes of oscillation frequencies and amplitudes have been measured
with the {\it Kepler} mission in the subdwarf B star KIC 10139564 by
\citep{2016A&A...594A..46Z}. Since these frequency and amplitude changes occur
with timescales that are several orders of magnitude shorter than
the cooling timescale of a DB white dwarf, these modulations of frequencies and
amplitudes have nothing to do with any evolutionary effect, such as,
e.g., neutrino cooling. Also, it is not expected that these
modulations are provoked by orbiting companions around the star, since
different timescales for the different triplets are involved. 
A clue to clarify this behavior could be a the possible interaction between
the components of the rotationally split triplets, that is, the presence
of nonlinear resonant mode coupling \citep[see, for details,][]
{1997A&A...321..159B}. If true, this star could constitute an excellent 
opportunity to study nonlinear effects in pulsating white dwarfs.  Note
that, unfortunately, the presence of frequency modulations
can make the measurement of the evolutionary (cooling) rate
of period change of KIC 08626021 unrealizable, thus  hampering any possibility of
placing constraints on the plasmon-neutrino emission rate.

\subsection{A V777 Her star in the field of the {\it K2} mission: PG 0112+104}

\cite{2017ApJ...835..277H} presented a pulsational analysis of the
already known DBV star PG 0112+104. With $T_{\rm eff} \gtrsim 30\,
000$ K,  this star is the hottest V777 Her star known, that defines
the blue edge of the DBV instability strip.  The star was analyzed on
the basis of $78.7$ days of nearly uninterrupted photometry from the
{\it Kepler} space telescope ---specifically the campaign 8 of the {\it K2}
extended mission. \cite{2017ApJ...835..277H} discovered 9 additional
periods apart from the 2 periods that were already known from
ground-based observations. The pulsation spectrum of this star
includes clear patterns of rotational splittings from consecutive
sequences of $\ell= 1$ and $\ell= 2$ modes. In addition, a surface rotational
period of 10.17 hr has been measured using an apparent spot
modulation.  By summarizing, this hot star is a promising candidate to
derive the dependence of rotational angular
velocity  with depth, that is, the differential rotation of the star,
through asteroseismology.
The estimation of the surface rotation independently from the spot
would constitute a strong test for the asteroseismological analysis.  On
the other hand, being PG 0112+104 such a hot DBV star, it constitutes
an excellent laboratory for studying plasmon neutrino production if
the rate of  period change for any mode is measured. Since the
pulsation frequencies in PG 0112+104 are extremely stable in phase,
the secular change of the periods could be detected and measured with
additional ---ground- or space-based--- observations. 

\subsection{ZZ Ceti stars in the nominal {\it Kepler} mission field}
\label{nominal-zzceti}

\cite{2011ApJ...741L..16H} reported the discovery of the first
identified DAV, WD J1916+3938 (Kepler ID 4552982),
in the original field of the {\it Kepler} mission. This ZZ Ceti star was
first identified through ground-based, time-series
photometry. A follow-up spectroscopic analysis indicated that it is a
DA white dwarf with $T_{\rm eff}= 11\,129 \pm 115$ K and $\log g = 8.34 \pm 0.06$,
 situating it near the hot boundary of the ZZ Ceti instability domain.
The object showed up to 0.5 \% amplitude variability at several
periods between 800 and 1450 s \citep{2015ApJ...809...14B}.
This star was submitted for
{\it Kepler} short-cadence observations. These observations led to
the unexpected discovery of a new phenomenon: outburst-like events in
DAV stars with effective temperatures near the  cool edge of the
instability strip \citep[][see Sect. \ref{outbursting} for details]{2015ApJ...809...14B}.
 
The second DAV in the {\it Kepler} field, KIC 11911480,
was discovered by \cite{2014MNRAS.438.3086G} using the
{\it Kepler}-INT Survey \citep[KIS;][]{2012AJ....144...24G}
to select white-dwarf candidates with colour-colours diagrams.  The variable nature
of KIC 11911480 was confirmed using ground-based time series photometry.
This star is close to the blue edge of the ZZ Ceti instability strip
($T_{\rm eff}= 12\,160 \pm 250$ K and $\log g= 7.94 \pm 0.10$).
The star was scrutinized along six months with the short-cadence mode of the
{\it Kepler} telescope, and a total of six independent
pulsation periods in the range $172.9-324.5$ s
---typical of the hot ZZ Ceti stars--- were detected. A preliminary
analysis indicated that the star is rotating with a period of $\sim 3.5$
days. \cite{2015ASPC..493..169G} and \cite{2016MNRAS.457.2855G}
reported the pulsation properties of 3
additional DAV stars observed with {\it Kepler}: KIC 10132702
($T_{\rm eff}= 11048\pm217$ K, $\log g= 8.07\pm0.08$),
 KIC 04357037 ($T_{\rm eff}= 11898\pm200$ K, $\log g= 8.03\pm0.08$), and
 KIC 07594781 ($T_{\rm eff}= 12217\pm1700$ K, $\log g= 7.54\pm0.22$). 
 Using the frequency spacing in the triplest and doublets, these authors
 estimated the rotation periods of these stars, obtaining values in the
 range ($0.9-3.2$) days. 

\subsection{ZZ Ceti stars monitored with the {\it K2}  mission} 
 
The first DAV star observed with the {\it K2} mission was GD 1212 \citep{2014ApJ...789...85H}.
This  star, which was already known to be pulsating  \citep{2006AJ....132..831G},
is a cool DAV with $T_{\rm eff}= 10\,970\pm170$ K and $\log g= 8.03\pm0.05$,
corresponding to a mass of $0.62\pm0.03M_{\odot}$. {\it K2} short-cadence observations revealed
at least 19 independent pulsation modes, ranging from $828.2$ s to $1220.8$ s,
and at least 17 nonlinear combination frequencies. This star also
exhibits  amplitude and frequency variations on timescales less than a week, 
reminiscent to the behavior detected by \cite{2016A&A...585A..22Z}
in the DBV star KIC 08626021 (see at the end of Sect. \ref{KIC8626021}).
This phenomenon  makes a precise determination of the pulsation periods of 
GD 1212 more complicated. Since this star is a cool ZZ Ceti, it would be expected that it were
characterized by high amplitudes. However,  the independent modes detected,
as well as the frequency combinations, show low amplitudes.

Another DAV star analyzed with the {\it K2} mission is SDSS J113655.17+040952.6
(hereafter SDSS J1136+0409), the first known pulsating DA white-dwarf in a post-common
envelope binary system with
a main-sequence (dM) companion star \citep{2015MNRAS.447..691P}. The ZZ Ceti
component has $T_{\rm eff}= 12\,330 \pm260$ K and $M_{\star}= 0.601\pm0.036 M_{\odot}$.
Nearly 78 days of {\it K2} observations allowed \cite{2015MNRAS.451.1701H} to detect seven
independent pulsation modes, three of which are rotationally split multiplets
compatible with a rotation period of $2.49\pm0.53$ h.

Usually, observations of pulsating white dwarfs with the {\it Kepler} space telescope 
have to be performed using the short-cadence mode of observations (1-min exposures)
to sufficiently over-sample typical white-dwarf pulsation periods (3-20 min)
for straightforward frequency measurement. However, \cite{2017ApJ...851...24B}
have demonstrated that it is possible to combine long-cadence {\it K2} data (30 min exposures)
with high-speed follow-up ground-based observations, to derive accurate pulsation
periods to a precision of $\sim 0.01 \mu$Hz. Using this approach, \cite{2017ApJ...851...24B}
have discovered two new ZZ Ceti variables from {\it K2} long-cadence data:
EPIC 210377280 and EPIC 220274129. For EPIC 220274129, \cite{2017ApJ...851...24B} inferred a stellar
rotation period of $12.7\pm1.3$ hr.

\begin{figure}[h!]
\begin{center}
\includegraphics[width=17cm]{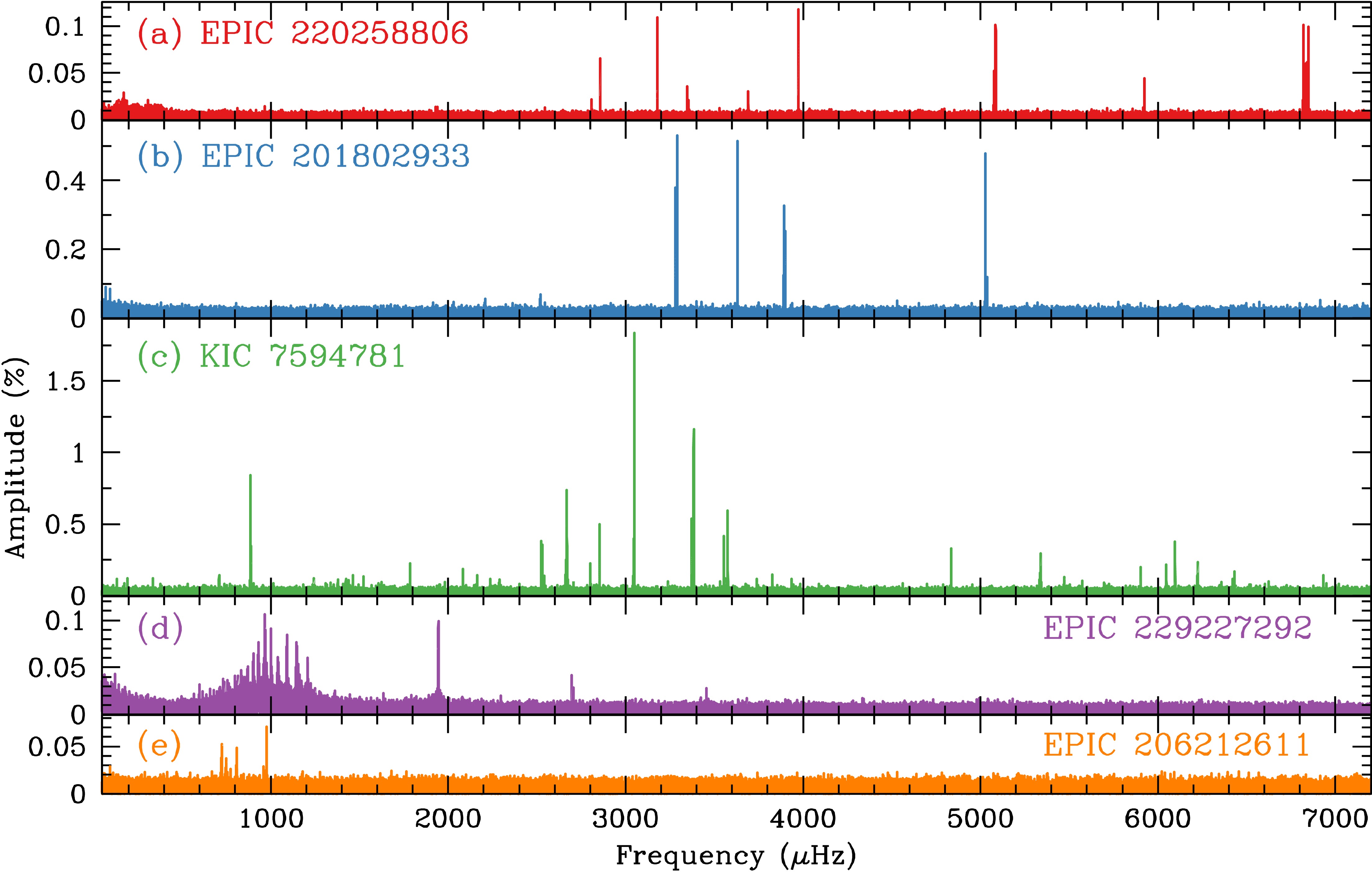}
\end{center}
\caption{The Fourier Transform corresponding to five representative ZZ Ceti stars
  observed with {\it K2}, with decreasing $T_{\rm eff}$ from top to bottom panel
  \citep{2017ApJS..232...23H}. The figures highlight the different phases of DAV stars along
  the ZZ Ceti instability strip. Reproduced by permission of the AAS.}
   \label{fig:5}
\end{figure}

The most complete and detailed study of ZZ Ceti stars observed
with the {\it Kepler} space telescope has been presented by \cite{2017ApJS..232...23H}.
This monumental work, based on the observations of 27 ZZ Ceti stars
collected with the nominal {\it Kepler} mission and mostly with the extended
{\it K2} mission, presents in detail the occurrence of outbursts in ZZ Ceti stars
near the red edge of the DAV instability strip (see Sect. \ref{outbursting}),
the discovery of a dichotomy of the widths of peaks in the Fourier spectrum (Sect. \ref{dichotomy}),
and a possible relationship between the rotation periods derived from
frequency splittings for 20 of the 27 DAVs and their stellar mass (see Sect. \ref{rotation}).
In addition, \cite{2017ApJS..232...23H} have been able to characterize
the DAV instability strip as observed by {\it Kepler} and {\it K2}. Indeed,
they show how the different trends regarding the length of the excited periods
and the amplitude of the modes evolve as the effective temperature decreases
along the instability strip. In particular, \cite{2017ApJS..232...23H} find
that the excited periods are short and with low amplitude near the blue edge
of the instability domain, but then the excited periods grow
(modes of larger radial orders are excited) and also the amplitude increases
with decreasing $T_{\rm eff}$. Towards effective temperatures near the cool boundary
  of the instability strip,
some ZZ Ceti show outbursts. Finally, at the red edge, the excited modes of DAVs are
the longest-period pulsations with relatively small amplitude. Fig. \ref{fig:5},
that displays the Fourier spectrum of ZZ Ceti stars at different effective temperatures, 
dramatically illustrates the described trend.

\subsection{Outbursting ZZ Ceti stars}
\label{outbursting}

\begin{figure}[h!]
\begin{center}
\includegraphics[width=17cm]{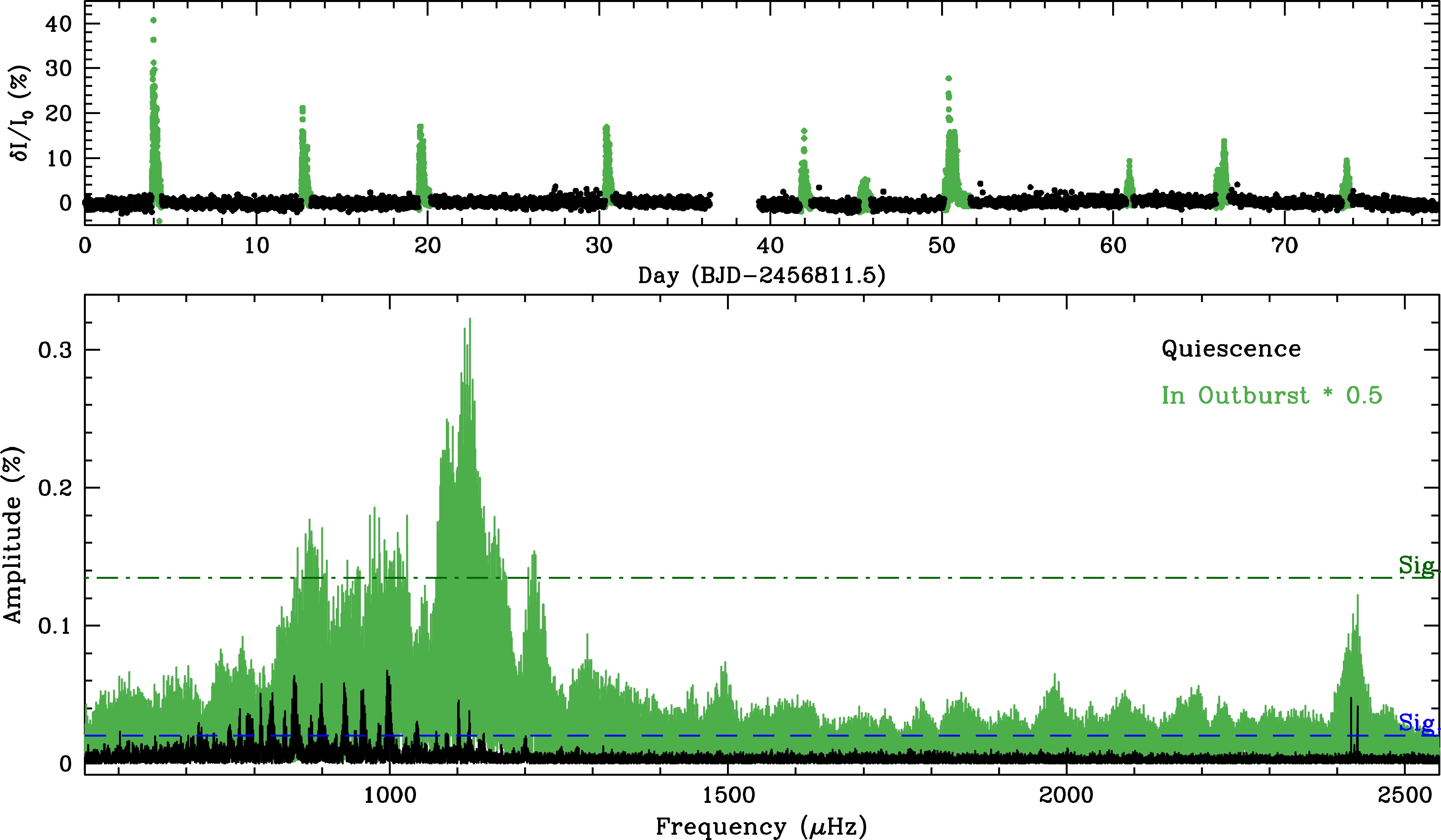}
\end{center}
\caption{Upper panel: the complete {\it K2} light curve of PG 1149+057 derived
  by \cite{2015ApJ...810L...5H}. The 10 outburst
  events are emphasized with green dots and the data in quiescence are marked with
  black dots. Bottom panel: the Fourier Transforms during and out of outburst (green and black).
  In both cases, pulsations have amplitudes larger than the corresponding
  significance thresholds, shown as dark green dashed–dotted and blue dashed lines.
  Note that normal-mode pulsations remain during the outbursts, but they have
  larger amplitudes and higher frequencies than in quiescence.
  Reproduced by permission of the AAS.}
   \label{fig:6}
\end{figure}

 The {\it Kepler} spacecraft observations of ZZ Ceti stars led to the discovery of a
 energetic phenomenon never observed before from the ground
  in this type of quiet pulsating stars: outburst-like events \citep{2017ASPC..509..303B}.
So far, this constitutes one of the most transcendent discoveries
(perhaps {\it the most} transcendent) in relation to pulsations in white dwarfs
observed with {\it Kepler}. An important aspect of these events
is that they repeat chaotically, with no fixed period of recurrence.
The first outbursting DAV,
KIC4552982\footnote{This star was the first ZZ Ceti identified in
  the original {\it Kepler} mission field
  \citep[][; see Sect. \ref{nominal-zzceti}]{2011ApJ...741L..16H}.},
which is located near the red edge of the ZZ Ceti instability strip
($T_{\rm eff}= 11\,129 \pm 115$ K, $\log= 8.34 \pm 0.06$),
was observed during $\sim 20$ months
with the nominal {\it Kepler} mission, yielding the 
longest pseudo-continuous light curve of a ZZ Ceti ever obtained
\citep{2015ApJ...809...14B}.  These observations allowed the detection of
20 pulsation modes with periodicities commonly observed in  ZZ Ceti stars,
along with 178 increments of brightness typical of outburst
phenomena, with peaks of up to 17 \% above the quiescent level.
The outbursts involve very energetic events ($E\sim 10^{33}$ erg), with a mean
recurrence period of about 2.7 days  and
4–25 hr of duration \citep{2015ApJ...809...14B}.  The second outbursting star, the already
known cool DAV PG 1149+057 ($T_{\rm eff}= 11\,060 \pm 170$ K, $\log= 8.06 \pm 0.05$),
was analyzed by \cite{2015ApJ...810L...5H}
with nearly continuous {\it K2} observations for more than 78.8 days.
This star shows outbursts with a typical recurrence time of $\sim 8$ day and 15 hr
of duration, and very large flux enhancements of up to 45\% above the
quiescent level.  For this star, the outbursts have a measurable impact on
  the spectrum of $g$ modes (in amplitude and
  frequency), demonstrating  that \emph{outbursts are an intrinsic phenomenon of the
    (otherwise isolated) star!} In Fig. \ref{fig:6} we show the complete {\it K2} light
curve of PG 1149+057 (upper panel)
and the Fourier Transforms for the case during and out of outburst
(lower panels). At the time of writing this article, a total of 8
outbursting (isolated) DAV stars have been discovered
\citep{2016ApJ...829...82B,2017ASPC..509..303B}, all of them
populating the red edge of the ZZ Ceti instability strip
\citep[see Fig. 12 of][]{2019A&ARv..27....7C}. This
strongly suggests that outbursts could be related to the origin
of the cool edge of the DAV instability domain, 
that is, the cessation of the $g$-mode pulsations \citep{2017ASPC..509..303B}.
A possible explanation for both the occurrence of outbursts and the origin of the
cool edge of the ZZ Ceti instability strip, has been conceived by \cite{2018ApJ...863...82L}.
It is connected with parametric instability through mode coupling of white-dwarf pulsations
\citep{1982AcA....32..147D,2001ApJ...546..469W,2018ApJ...863...82L}.

\begin{figure}[h!]
\begin{center}
\includegraphics[width=17cm]{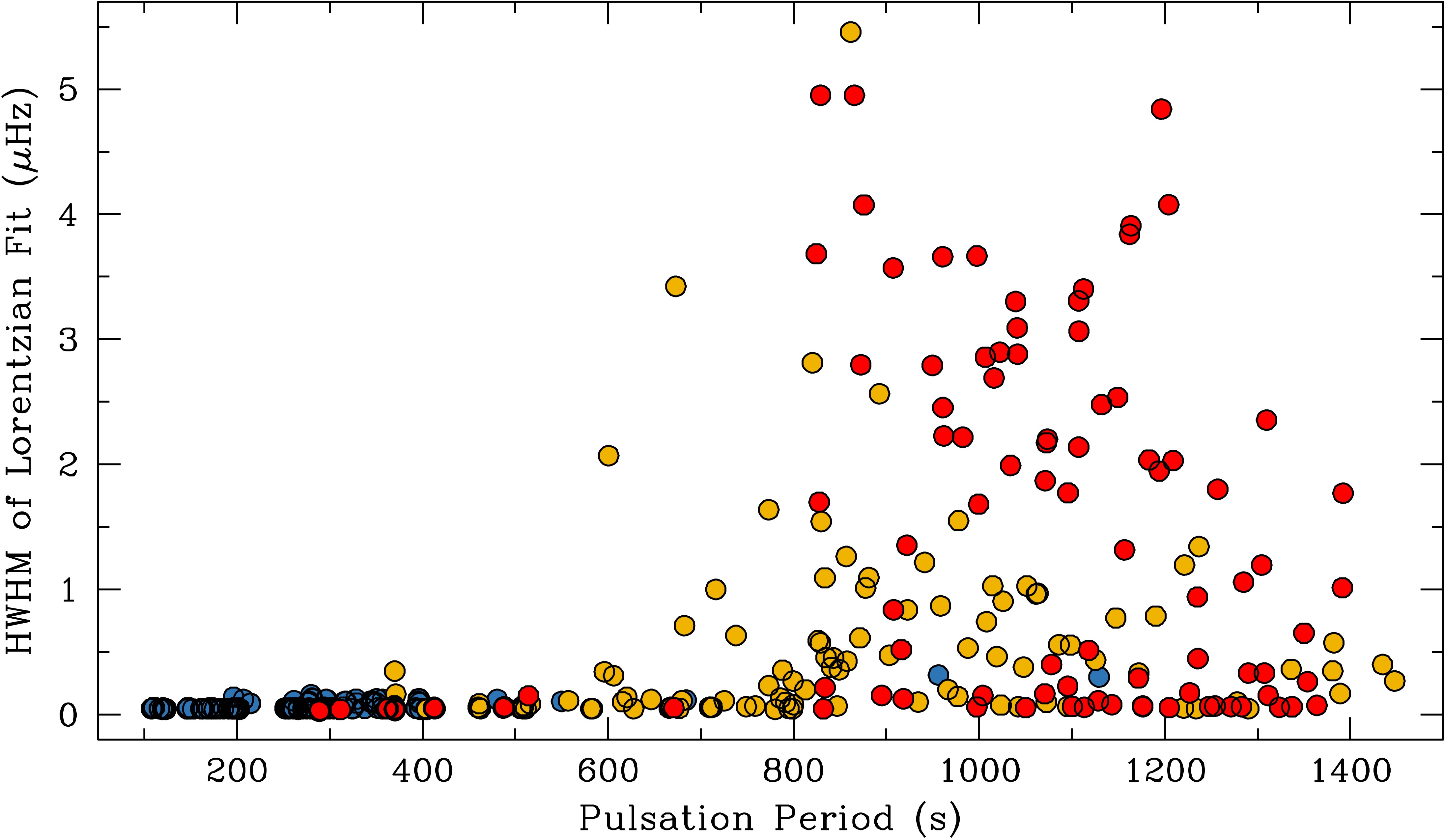}
\end{center}
\caption{Half-width at half-maximum (HWHM) of Lorentzian functions fit to all
  significant peaks (corresponding to genuine eigenmodes) in the power spectra of
   27 DAVs observed through K2 \citep{2017ApJS..232...23H}. Blue circles are for
  objects with weighted mean period (WMP) $ < 600$ s, gold with WMP $> 600$ s,
  and red those with outbursts.
  A clear increase in HWHM at $\sim 800$ s is present. Reproduced by permission of the AAS.}
   \label{fig:7}
\end{figure}

\subsection{Dichotomy of mode-line widths in ZZ Ceti stars}
\label{dichotomy}

Another major observational result from the {\it Kepler} space telescope
is the discovery by \cite{2017ApJS..232...23H} of a clear dichotomy of oscillation
mode-line widths in the power spectrum of 27 DAVs  observed from
space.  Indeed, $g$ modes characterized by periods longer than
roughly $800$ s are generally incoherent (large line widths) over the length of
observations, while $g$ modes with periods $\lesssim 800$ s are observed to
be much more stable in phase and amplitude (smaller line width)\footnote{This
  phenomenon cannot be associated to stochastic mode driving, that for white dwarfs would
excite pulsations with very short periods, of the order of $\sim 1$
sec \citep{2013EPJWC..4305005S}.}. We depict in Fig. \ref{fig:7} \citep[extracted from][]{2017ApJS..232...23H} the
half-width at half-maximum (HWHM) of  all significant peaks
(excluding any nonlinear combination frequencies) in the power spectra of
 27 DAVs observed through {\it Kepler} and {\it K2}.
 The dichotomy of  mode-line widths could be intimately linked to 
the oscillation of the outer convection zone of a DA white dwarf
during pulsations. Specifically, the oscillation out and in of the base of the 
convection zone would affect the radial eigenfunction of certain $g$ modes --- those
having the outer turning point of oscillation located precisely at the base of the
convection zone \citep{2020ApJ...890...11M}.
This mechanism may be relevant for the limitation of pulsation amplitudes in
pulsating white dwarfs for modes with periods above a threshold period.

\subsection{White-dwarf rotation rates from asteroseismology of ZZ Ceti stars}
\label{rotation}

 \cite{2017ApJS..232...23H} have also derived the rotation rates of 
20 isolated DAVs observed with {\it Kepler} and {\it K2} using
rotational splittings, this way doubling the number of pulsating white dwarfs
with measured rotation rates via asteroseismology\footnote{Note that rotation rates can sometimes
  also be measured from long-cadence data, as in \cite{2017ApJ...851...24B}.
\citep[see Table \ref{table:2} of this article and
  Table 10 of][]{2019A&ARv..27....7C}. The results indicate that the average
rotation period is of $\sim 30-40$ hours, although ZZ Ceti stars that rotate
faster (with periods of $\sim 1-2$ hours) have also been found.} There is
some evidence of the existence of a correlation between high mass and fast
rotation, an idea reinforced by the fastest rotation ever measured in
an isolated white dwarf\footnote{Except for
  SDSSJ125230.93$-$023417.72, which is the fastest-rotating apparently
  isolated  white dwarf yet discovered, that
  rotates with a period of $317.278\pm0.013$ seconds \citep{2020ApJ...894...19R}.},
of 1.13 hour, for the massive ZZ Ceti star
SDSSJ0837+1856 ($M_{\star}= 0.87 M_{\odot}$) by
\cite{2017ApJ...841L...2H},  although more high-mass white dwarfs are
necessary to confirm this trend.  \cite{2017ApJS..232...23H} find that
most of isolated  descendants of $1.7–3.0 M_{\odot}$ ZAMS
progenitors rotate at $\sim 1.5$ days, instead of minutes, what would
be expected if the angular momentum of a $3.0 M_{\odot}$ MS star with
an initial rotation period of 10 hr, were fully conserved
\citep{2004IAUS..215..561K}.  This fact strongly indicates that most
internal angular momentum must be lost on the first-ascent giant
branch.

\subsection{Asteroseismology of {\it Kepler} ZZ Ceti stars
  with fully evolutionary models}
\label{astero}

In Sect. \ref{KIC8626021} we described in detail the various
asteroseismological analyses conducted to elucidate the internal
structure of the {\it Kepler} DBV star KIC 8626021. Here, we briefly
describe the asteroseismological analysis on ZZ Cetis observed with {\it Kepler}
carried out by \cite{2017ApJ...851...60R}. This analysis was based on evolutionary
and pulsational white-dwarf models computed with the {\tt LPCODE} and {\tt LP-PUL}
codes, adopting the La Plata Group's approach
of white-dwarf asteroseismology (see at the end of Sect. \ref{pulsation_properties_classes}).
Specifically, the first four published ZZ Ceti stars observed
with the {\it Kepler} mission, that is, EPIC 60017836 (GD 1212), EPIC 201730811
(SDSS J113655.17+040952.6), KIC 11911480, and KIC 4552982, were studied. Currently,
this remains the only detailed asteroseismological study of DAVs
observed with the {\it Kepler} space telescope.
A period-to-period fit analysis of the four target stars was carried out by employing a grid of full
evolutionary  models representative  of C/O-core DA  white-dwarf stars
constructed with detailed and updated input physics. These models have
consistent chemical profiles for both the core and the envelope for
various stellar masses, particularly suitable for 
asteroseismological fits of ZZ Ceti stars.  The chemical profiles of
the models were computed considering the complete evolution of the
progenitor stars from the ZAMS through the thermally pulsing and
mass-loss phases on the asymptotic giant  branch (AGB).

 Apart from the observed periods, the analysis by
\cite{2017ApJ...851...60R} incorporated the amplitude of the modes,
rotational splitting multiplets, and period spacing data, as well as photometry and
spectroscopy information. For each analyzed star, an asteroseismological model
that closely reproduces observed properties such as periods, stellar mass
and effective temperature was presented. In
the particular case of KIC 11911480 and EPIC 201730811, the
asteroseismological masses are similar each other, although the H
envelope for EPIC 201730811 (that is part of a binary system and likely went
through a common envelope phase) is 10 times thinner than for  KIC
11911480. This result confirms what was discussed in
\cite{2015MNRAS.451.1701H}  on the basis of a preliminary
  asteroseismological analysis,
that the H-layer mass of  EPIC 201730811 is roughly $10^{-5} M_{\star}$. In the case of
KIC 4552982, which is a red-edge ZZ Ceti
star, the asteroseismological model has a very thin H envelope mass
($M_{\rm H}/M_{\odot}= 4.7 \times 10^{-9}$), which could be related to
the outburst nature of this star, as reported by
\cite{2015ApJ...809...14B}. Whether this is a common characteristic
between all the outbursting DAVs or not is a very interesting topic
that deserves to be explored. Finally, in the case of EPIC 60017836
(other red-edge DAV), an asteroseismological model was obtained having
a stellar mass compatible with the atmospheric parameters from
photometry combined with parallax and spectroscopy.

\section{Conclusions}

The {\it Kepler} space telescope --- both the nominal {\it Kepler}
mission and the extended mission {\it K2} ---  has been very
successful in the search for extrasolar planets and has also meant a
revolution in the area of asteroseismology for many classes of
pulsating stars.  In the particular case of pulsating white dwarfs,
the numbers speak for themselves. In total, $2\,166$ white dwarfs were
observed by {\it Kepler}/{\it K2}. There were 81 DAVs observed by {\it
  Kepler}/{\it K2}, with 75 of them getting short cadence data. Of
them, the analyses of 32 DAVs have been published (29 stars in
journals and 3 objects analyzed in a PhD Thesis; see Table
\ref{table:2}).  Thanks to {\it Kepler}/{\it K2}, the number of ZZ
Ceti stars with rotation period measured with asteroseismology has
increased by a factor of 2. Regarding DBVs, 7 stars were observed,
although the analysis of only 2 of them have been published. Finally,
2 unique GW Vir  stars were observed, including the prototypical star
PG 1159$-$035, but no results have been published yet.

The story of the {\it Kepler} mission and the pulsating white dwarf
stars does not end here.  There is an immense amount of data waiting
to be analyzed, and this will likely lead to new discoveries in the
near future. In turn, this will drive new asteroseismological
analyzes, in addition to those already published. All in all, the {\it
  Kepler} space telescope has had a strong impact (and will continue
to have it) in the area of white-dwarf asteroseismology.  This first
step will be multiplied by current and future space missions, such as
the Transiting Exoplanet Survey Satellite \citep[{\it TESS};][]{2015JATIS...1a4003R},
which is already in full operation
and providing the first results on pulsating white dwarfs
\citep{2019A&A...632A..42B,2020A&A...633A..20A,2020arXiv200311481B},
and other space missions that will begin to be operational in the coming years, such
as Cheops \citep{2018A&A...620A.203M} and Plato
\citep{2018EPSC...12..969P}. 

\section*{Acknowledgments}

 I thank the anonymous referees for their valuable suggestions  that  improved
the  content  and  presentation  of  the  paper.
I warmly thank Leandro G. Althaus, Keaton J. Bell and J. J. Hermes
for being so kind to read the manuscript and suggest relevant changes that greatly
improved its content. Part of this work was supported by
AGENCIA through the Programa  de  Modernizaci\'on  Tecnol\'ogica  BID  1728/OC-AR,
by  the  PIP  112-200801-00940 grant from CONICET, and  by the grant  G149  from
University  of  La  Plata. This research has made intensive use of NASA Astrophysics
Data System.

\bibliographystyle{frontiersinSCNS_ENG_HUMS}
\bibliography{corsico_bibliografia}

\begin{thebibliography}{137}
\providecommand{\natexlab}[1]{#1}
\expandafter\ifx\csname urlstyle\endcsname\relax
  \providecommand{\doi}[1]{doi:\discretionary{}{}{}#1}\else
  \providecommand{\doi}{doi:\discretionary{}{}{}\begingroup
  \urlstyle{rm}\Url}\fi
\providecommand{\selectlanguage}[1]{\relax}
\providecommand{\bibAnnoteFile}[1]{%
  \IfFileExists{#1}{\begin{quotation}\noindent\textsc{Key:} #1\\
  \textsc{Annotation:}\ \input{#1}\end{quotation}}{}}
\providecommand{\bibAnnote}[2]{%
  \begin{quotation}\noindent\textsc{Key:} #1\\
  \textsc{Annotation:}\ #2\end{quotation}}

\bibitem[{{Aerts} et~al.(2010){Aerts}, {Christensen-Dalsgaard}, and
  {Kurtz}}]{2010aste.book.....A}
{Aerts}, C., {Christensen-Dalsgaard}, J., and {Kurtz}, D.~W. (2010).
\newblock \emph{{Asteroseismology}}
\bibAnnoteFile{2010aste.book.....A}

\bibitem[{{Althaus} et~al.(2010){Althaus}, {C{\'o}rsico}, {Isern}, and
  {Garc{\'{\i}}a-Berro}}]{2010A&ARv..18..471A}
{Althaus}, L.~G., {C{\'o}rsico}, A.~H., {Isern}, J., and {Garc{\'{\i}}a-Berro},
  E. (2010).
\newblock {Evolutionary and pulsational properties of white dwarf stars}.
\newblock \emph{\aapr} 18, 471--566.
\newblock \doi{10.1007/s00159-010-0033-1}
\bibAnnoteFile{2010A&ARv..18..471A}

\bibitem[{{Althaus} et~al.(2020){Althaus}, {C{\'o}rsico}, {Uzundag},
  {Vu{\v{c}}kovi{\'c}}, {Baran}, {Bell} et~al.}]{2020A&A...633A..20A}
{Althaus}, L.~G., {C{\'o}rsico}, A.~H., {Uzundag}, M., {Vu{\v{c}}kovi{\'c}},
  M., {Baran}, A.~S., {Bell}, K.~J., et~al. (2020).
\newblock {About the existence of warm H-rich pulsating white dwarfs}.
\newblock \emph{\aap} 633, A20.
\newblock \doi{10.1051/0004-6361/201936346}
\bibAnnoteFile{2020A&A...633A..20A}

\bibitem[{{Althaus} et~al.(2013){Althaus}, {Miller Bertolami}, and
  {C{\'o}rsico}}]{2013A&A...557A..19A}
{Althaus}, L.~G., {Miller Bertolami}, M.~M., and {C{\'o}rsico}, A.~H. (2013).
\newblock {New evolutionary sequences for extremely low-mass white dwarfs.
  Homogeneous mass and age determinations and asteroseismic prospects}.
\newblock \emph{\aap} 557, A19.
\newblock \doi{10.1051/0004-6361/201321868}
\bibAnnoteFile{2013A&A...557A..19A}

\bibitem[{{Althaus} et~al.(2005){Althaus}, {Miller Bertolami}, {C{\'o}rsico},
  {Garc{\'{\i}}a-Berro}, and {Gil-Pons}}]{2005A&A...440L...1A}
{Althaus}, L.~G., {Miller Bertolami}, M.~M., {C{\'o}rsico}, A.~H.,
  {Garc{\'{\i}}a-Berro}, E., and {Gil-Pons}, P. (2005).
\newblock {The formation of DA white dwarfs with thin hydrogen envelopes}.
\newblock \emph{\aap} 440, L1--L4.
\newblock \doi{10.1051/0004-6361:200500159}
\bibAnnoteFile{2005A&A...440L...1A}

\bibitem[{{Bedin} et~al.(2015){Bedin}, {Salaris}, {Anderson}, {Cassisi},
  {Milone}, {Piotto} et~al.}]{2015MNRAS.448.1779B}
{Bedin}, L.~R., {Salaris}, M., {Anderson}, J., {Cassisi}, S., {Milone}, A.~P.,
  {Piotto}, G., et~al. (2015).
\newblock {Hubble Space Telescope observations of the Kepler-field cluster NGC
  6819 - I. The bottom of the white dwarf cooling sequence}.
\newblock \emph{mnras} 448, 1779--1788.
\newblock \doi{10.1093/mnras/stv069}
\bibAnnoteFile{2015MNRAS.448.1779B}

\bibitem[{{Bedin} et~al.(2009){Bedin}, {Salaris}, {Piotto}, {Anderson}, {King},
  and {Cassisi}}]{2009ApJ...697..965B}
{Bedin}, L.~R., {Salaris}, M., {Piotto}, G., {Anderson}, J., {King}, I.~R., and
  {Cassisi}, S. (2009).
\newblock {The End of the White Dwarf Cooling Sequence in M4: An Efficient
  Approach}.
\newblock \emph{apj} 697, 965--979.
\newblock \doi{10.1088/0004-637X/697/2/965}
\bibAnnoteFile{2009ApJ...697..965B}

\bibitem[{Bell(2017)}]{2017PhDT........14C}
Bell, K.~J. (2017).
\newblock \emph{{Pulsational oddities at the extremes of the DA white dwarf
  instability strip}}.
\newblock Ph.D. thesis, University of Texas
\bibAnnoteFile{2017PhDT........14C}

\bibitem[{{Bell} et~al.(2019){Bell}, {C{\'o}rsico}, {Bischoff-Kim}, {Althaus},
  {Bradley}, {Calcaferro} et~al.}]{2019A&A...632A..42B}
{Bell}, K.~J., {C{\'o}rsico}, A.~H., {Bischoff-Kim}, A., {Althaus}, L.~G.,
  {Bradley}, P.~A., {Calcaferro}, L.~M., et~al. (2019).
\newblock {TESS first look at evolved compact pulsators. Asteroseismology of
  the pulsating helium-atmosphere white dwarf TIC 257459955}.
\newblock \emph{\aap} 632, A42.
\newblock \doi{10.1051/0004-6361/201936340}
\bibAnnoteFile{2019A&A...632A..42B}

\bibitem[{{Bell} et~al.(2015){Bell}, {Hermes}, {Bischoff-Kim}, {Moorhead},
  {Montgomery}, {{\O}stensen} et~al.}]{2015ApJ...809...14B}
{Bell}, K.~J., {Hermes}, J.~J., {Bischoff-Kim}, A., {Moorhead}, S.,
  {Montgomery}, M.~H., {{\O}stensen}, R., et~al. (2015).
\newblock {KIC 4552982: Outbursts and Asteroseismology from the Longest
  Pseudo-continuous Light Curve of a ZZ Ceti}.
\newblock \emph{\apj} 809, 14.
\newblock \doi{10.1088/0004-637X/809/1/14}
\bibAnnoteFile{2015ApJ...809...14B}

\bibitem[{{Bell} et~al.(2016){Bell}, {Hermes}, {Montgomery}, {Gentile Fusillo},
  {Raddi}, {G{\"a}nsicke} et~al.}]{2016ApJ...829...82B}
{Bell}, K.~J., {Hermes}, J.~J., {Montgomery}, M.~H., {Gentile Fusillo}, N.~P.,
  {Raddi}, R., {G{\"a}nsicke}, B.~T., et~al. (2016).
\newblock {Outbursts in Two New Cool Pulsating DA White Dwarfs}.
\newblock \emph{\apj} 829, 82.
\newblock \doi{10.3847/0004-637X/829/2/82}
\bibAnnoteFile{2016ApJ...829...82B}

\bibitem[{{Bell} et~al.(2017{\natexlab{a}}){Bell}, {Hermes}, {Montgomery},
  {Winget}, {Gentile Fusillo}, {Raddi} et~al.}]{2017ASPC..509..303B}
{Bell}, K.~J., {Hermes}, J.~J., {Montgomery}, M.~H., {Winget}, D.~E., {Gentile
  Fusillo}, N.~P., {Raddi}, R., et~al. (2017{\natexlab{a}}).
\newblock {The First Six Outbursting Cool DA White Dwarf Pulsators}.
\newblock In \emph{20th European White Dwarf Workshop}, eds. P.-E. {Tremblay},
  B.~{Gaensicke}, and T.~{Marsh}. vol. 509 of \emph{Astronomical Society of the
  Pacific Conference Series}, 303
\bibAnnoteFile{2017ASPC..509..303B}

\bibitem[{{Bell} et~al.(2017{\natexlab{b}}){Bell}, {Hermes}, {Vanderbosch},
  {Montgomery}, {Winget}, {Dennihy} et~al.}]{2017ApJ...851...24B}
{Bell}, K.~J., {Hermes}, J.~J., {Vanderbosch}, Z., {Montgomery}, M.~H.,
  {Winget}, D.~E., {Dennihy}, E., et~al. (2017{\natexlab{b}}).
\newblock {Destroying Aliases from the Ground and Space: Super-Nyquist ZZ Cetis
  in K2 Long Cadence Data}.
\newblock \emph{\apj} 851, 24.
\newblock \doi{10.3847/1538-4357/aa9702}
\bibAnnoteFile{2017ApJ...851...24B}

\bibitem[{{Bergeron} et~al.(2011){Bergeron}, {Wesemael}, {Dufour}, {Beauchamp},
  {Hunter}, {Saffer} et~al.}]{2011ApJ...737...28B}
{Bergeron}, P., {Wesemael}, F., {Dufour}, P., {Beauchamp}, A., {Hunter}, C.,
  {Saffer}, R.~A., et~al. (2011).
\newblock {A Comprehensive Spectroscopic Analysis of DB White Dwarfs}.
\newblock \emph{\apj} 737, 28.
\newblock \doi{10.1088/0004-637X/737/1/28}
\bibAnnoteFile{2011ApJ...737...28B}

\bibitem[{{Bessel}(1844)}]{1844MNRAS...6R.136B}
{Bessel}, F.~W. (1844).
\newblock {On the variations of the proper motions of Procyon and Sirius}.
\newblock \emph{\mnras} 6, 136--141.
\newblock \doi{10.1093/mnras/6.11.136}
\bibAnnoteFile{1844MNRAS...6R.136B}

\bibitem[{{Bischoff-Kim} and {{\O}stensen}(2011)}]{2011ApJ...742L..16B}
{Bischoff-Kim}, A. and {{\O}stensen}, R.~H. (2011).
\newblock {Asteroseismology of the Kepler Field DBV White Dwarf. It is a Hot
  One}.
\newblock \emph{\apjl} 742, L16.
\newblock \doi{10.1088/2041-8205/742/1/L16}
\bibAnnoteFile{2011ApJ...742L..16B}

\bibitem[{{Bischoff-Kim} et~al.(2014){Bischoff-Kim}, {{\O}stensen}, {Hermes},
  and {Provencal}}]{2014ApJ...794...39B}
{Bischoff-Kim}, A., {{\O}stensen}, R.~H., {Hermes}, J.~J., and {Provencal},
  J.~L. (2014).
\newblock {Seven-period Asteroseismic Fit of the Kepler DBV}.
\newblock \emph{\apj} 794, 39.
\newblock \doi{10.1088/0004-637X/794/1/39}
\bibAnnoteFile{2014ApJ...794...39B}

\bibitem[{{Bischoff-Kim} et~al.(2019){Bischoff-Kim}, {Provencal}, {Bradley},
  {Montgomery}, {Shipman}, {Harrold} et~al.}]{2019ApJ...871...13B}
{Bischoff-Kim}, A., {Provencal}, J.~L., {Bradley}, P.~A., {Montgomery}, M.~H.,
  {Shipman}, H.~L., {Harrold}, S.~T., et~al. (2019).
\newblock {GD358: Three Decades of Observations for the In-depth
  Asteroseismology of a DBV Star}.
\newblock \emph{\apj} 871, 13.
\newblock \doi{10.3847/1538-4357/aae2b1}
\bibAnnoteFile{2019ApJ...871...13B}

\bibitem[{{Bogn{\'a}r} et~al.(2020){Bogn{\'a}r}, {Kawaler}, {Bell}, {Schrand
  t}, {Baran}, {Bradley} et~al.}]{2020arXiv200311481B}
{Bogn{\'a}r}, Z., {Kawaler}, S.~D., {Bell}, K.~J., {Schrand t}, C., {Baran},
  A.~S., {Bradley}, P.~A., et~al. (2020).
\newblock {TESS first look at evolved compact pulsators: Known ZZ Ceti stars of
  the southern ecliptic hemisphere as seen by TESS}.
\newblock \emph{arXiv e-prints} , arXiv:2003.11481
\bibAnnoteFile{2020arXiv200311481B}

\bibitem[{{Borucki} et~al.(2010){Borucki}, {Koch}, {Basri}, {Batalha}, {Brown},
  {Caldwell} et~al.}]{2010Sci...327..977B}
{Borucki}, W.~J., {Koch}, D., {Basri}, G., {Batalha}, N., {Brown}, T.,
  {Caldwell}, D., et~al. (2010).
\newblock {Kepler Planet-Detection Mission: Introduction and First Results}.
\newblock \emph{Science} 327, 977.
\newblock \doi{10.1126/science.1185402}
\bibAnnoteFile{2010Sci...327..977B}

\bibitem[{{Brassard} et~al.(1991){Brassard}, {Fontaine}, {Wesemael}, {Kawaler},
  and {Tassoul}}]{1991ApJ...367..601B}
{Brassard}, P., {Fontaine}, G., {Wesemael}, F., {Kawaler}, S.~D., and
  {Tassoul}, M. (1991).
\newblock {Adiabatic properties of pulsating DA white dwarfs. I - The treatment
  of the Brunt-Vaisala frequency and the region of period formation}.
\newblock \emph{\apj} 367, 601--611.
\newblock \doi{10.1086/169655}
\bibAnnoteFile{1991ApJ...367..601B}

\bibitem[{{Brickhill}(1991)}]{1991MNRAS.251..673B}
{Brickhill}, A.~J. (1991).
\newblock {The pulsations of ZZ Ceti stars. III - The driving mechanism}.
\newblock \emph{\mnras} 251, 673--680.
\newblock \doi{10.1093/mnras/251.4.673}
\bibAnnoteFile{1991MNRAS.251..673B}

\bibitem[{{Brown} and {Gilliland}(1994)}]{1994ARA&A..32...37B}
{Brown}, T.~M. and {Gilliland}, R.~L. (1994).
\newblock {Asteroseismology}.
\newblock \emph{\araa} 32, 37--82.
\newblock \doi{10.1146/annurev.aa.32.090194.000345}
\bibAnnoteFile{1994ARA&A..32...37B}

\bibitem[{{Buchler} et~al.(1997){Buchler}, {Goupil}, and
  {Hansen}}]{1997A&A...321..159B}
{Buchler}, J.~R., {Goupil}, M.~J., and {Hansen}, C.~J. (1997).
\newblock {On the role of resonances in nonradial pulsators.}
\newblock \emph{\aap} 321, 159--176
\bibAnnoteFile{1997A&A...321..159B}

\bibitem[{{Calcaferro} et~al.(2017){Calcaferro}, {C{\'o}rsico}, {Camisassa},
  {Althaus}, and {Shibahashi}}]{2017EPJWC.15206012C}
{Calcaferro}, L.~M., {C{\'o}rsico}, A.~H., {Camisassa}, M.~E., {Althaus},
  L.~G., and {Shibahashi}, H. (2017).
\newblock {Pulsational instability of high-luminosity H-rich pre-white dwarf
  star}.
\newblock In \emph{European Physical Journal Web of Conferences}. vol. 152 of
  \emph{European Physical Journal Web of Conferences}, 06012.
\newblock \doi{10.1051/epjconf/201715206012}
\bibAnnoteFile{2017EPJWC.15206012C}

\bibitem[{{Camisassa} et~al.(2019){Camisassa}, {Althaus}, {C{\'o}rsico}, {De
  Ger{\'o}nimo}, {Miller Bertolami}, {Novarino} et~al.}]{2019A&A...625A..87C}
{Camisassa}, M.~E., {Althaus}, L.~G., {C{\'o}rsico}, A.~H., {De Ger{\'o}nimo},
  F.~C., {Miller Bertolami}, M.~M., {Novarino}, M.~L., et~al. (2019).
\newblock {The evolution of ultra-massive white dwarfs}.
\newblock \emph{\aap} 625, A87.
\newblock \doi{10.1051/0004-6361/201833822}
\bibAnnoteFile{2019A&A...625A..87C}

\bibitem[{{Camisassa} et~al.(2016){Camisassa}, {C{\'o}rsico}, {Althaus}, and
  {Shibahashi}}]{2016A&A...595A..45C}
{Camisassa}, M.~E., {C{\'o}rsico}, A.~H., {Althaus}, L.~G., and {Shibahashi},
  H. (2016).
\newblock {Pulsations powered by hydrogen shell burning in white dwarfs}.
\newblock \emph{\aap} 595, A45.
\newblock \doi{10.1051/0004-6361/201628857}
\bibAnnoteFile{2016A&A...595A..45C}

\bibitem[{{Campos} et~al.(2016){Campos}, {Bergeron}, {Romero}, {Kepler},
  {Ourique}, {Costa} et~al.}]{2016MNRAS.456.3729C}
{Campos}, F., {Bergeron}, P., {Romero}, A.~D., {Kepler}, S.~O., {Ourique}, G.,
  {Costa}, J.~E.~S., et~al. (2016).
\newblock {A comparative analysis of the observed white dwarf cooling sequence
  from globular clusters}.
\newblock \emph{\mnras} 456, 3729--3742.
\newblock \doi{10.1093/mnras/stv2911}
\bibAnnoteFile{2016MNRAS.456.3729C}

\bibitem[{{Campos} et~al.(2013){Campos}, {Kepler}, {Bonatto}, and
  {Ducati}}]{2013MNRAS.433..243C}
{Campos}, F., {Kepler}, S.~O., {Bonatto}, C., and {Ducati}, J.~R. (2013).
\newblock {Multichromatic colour-magnitude diagrams of the globular cluster NGC
  6366}.
\newblock \emph{\mnras} 433, 243--250.
\newblock \doi{10.1093/mnras/stt719}
\bibAnnoteFile{2013MNRAS.433..243C}

\bibitem[{{Catelan} and {Smith}(2015)}]{2015pust.book.....C}
{Catelan}, M. and {Smith}, H.~A. (2015).
\newblock \emph{{Pulsating Stars}}
\bibAnnoteFile{2015pust.book.....C}

\bibitem[{{Chandrasekhar}(1939)}]{1939isss.book.....C}
{Chandrasekhar}, S. (1939).
\newblock \emph{{An introduction to the study of stellar structure}}
\bibAnnoteFile{1939isss.book.....C}

\bibitem[{{Charpinet} et~al.(2019){Charpinet}, {Brassard}, {Giammichele}, and
  {Fontaine}}]{2019A&A...628L...2C}
{Charpinet}, S., {Brassard}, P., {Giammichele}, N., and {Fontaine}, G. (2019).
\newblock {Improved seismic model of the pulsating DB white dwarf KIC 08626021
  corrected from the effects of neutrino cooling}.
\newblock \emph{\aap} 628, L2.
\newblock \doi{10.1051/0004-6361/201935823}
\bibAnnoteFile{2019A&A...628L...2C}

\bibitem[{{C{\'o}rsico} and {Althaus}(2006)}]{2006A&A...454..863C}
{C{\'o}rsico}, A.~H. and {Althaus}, L.~G. (2006).
\newblock {Asteroseismic inferences on GW Virginis variable stars in the frame
  of new PG 1159 evolutionary models}.
\newblock \emph{\aap} 454, 863--881.
\newblock \doi{10.1051/0004-6361:20054199}
\bibAnnoteFile{2006A&A...454..863C}

\bibitem[{{C{\'o}rsico} and {Althaus}(2014)}]{2014ApJ...793L..17C}
{C{\'o}rsico}, A.~H. and {Althaus}, L.~G. (2014).
\newblock {Short-period g-mode Pulsations in Low-mass White Dwarfs Triggered by
  H-shell Burning}.
\newblock \emph{\apjl} 793, L17.
\newblock \doi{10.1088/2041-8205/793/1/L17}
\bibAnnoteFile{2014ApJ...793L..17C}

\bibitem[{{C{\'o}rsico} et~al.(2013){C{\'o}rsico}, {Althaus},
  {Garc{\'\i}a-Berro}, and {Romero}}]{2013JCAP...06..032C}
{C{\'o}rsico}, A.~H., {Althaus}, L.~G., {Garc{\'\i}a-Berro}, E., and {Romero},
  A.~D. (2013).
\newblock {An independent constraint on the secular rate of variation of the
  gravitational constant from pulsating white dwarfs}.
\newblock \emph{\jcap} 2013, 032.
\newblock \doi{10.1088/1475-7516/2013/06/032}
\bibAnnoteFile{2013JCAP...06..032C}

\bibitem[{{C{\'o}rsico} et~al.(2008){C{\'o}rsico}, {Althaus}, {Kepler},
  {Costa}, and {Miller Bertolami}}]{2008A&A...478..869C}
{C{\'o}rsico}, A.~H., {Althaus}, L.~G., {Kepler}, S.~O., {Costa}, J.~E.~S., and
  {Miller Bertolami}, M.~M. (2008).
\newblock {Asteroseismological measurements on PG 1159-035, the prototype of
  the GW Virginis variable stars}.
\newblock \emph{\aap} 478, 869--881.
\newblock \doi{10.1051/0004-6361:20078646}
\bibAnnoteFile{2008A&A...478..869C}

\bibitem[{{C{\'o}rsico} et~al.(2012{\natexlab{a}}){C{\'o}rsico}, {Althaus},
  {Miller Bertolami}, and {Bischoff-Kim}}]{2012A&A...541A..42C}
{C{\'o}rsico}, A.~H., {Althaus}, L.~G., {Miller Bertolami}, M.~M., and
  {Bischoff-Kim}, A. (2012{\natexlab{a}}).
\newblock {Asteroseismology of the Kepler V777 Herculis variable white dwarf
  with fully evolutionary models}.
\newblock \emph{\aap} 541, A42.
\newblock \doi{10.1051/0004-6361/201118736}
\bibAnnoteFile{2012A&A...541A..42C}

\bibitem[{{C{\'o}rsico} et~al.(2009){C{\'o}rsico}, {Althaus}, {Miller
  Bertolami}, {Gonz{\'a}lez P{\'e}rez}, and {Kepler}}]{2009ApJ...701.1008C}
{C{\'o}rsico}, A.~H., {Althaus}, L.~G., {Miller Bertolami}, M.~M.,
  {Gonz{\'a}lez P{\'e}rez}, J.~M., and {Kepler}, S.~O. (2009).
\newblock {On the Possible Existence of Short-Period g-Mode Instabilities
  Powered by Nuclear-Burning Shells in Post-Asymptotic Giant Branch H-Deficient
  (PG1159-Type) Stars}.
\newblock \emph{\apj} 701, 1008--1014.
\newblock \doi{10.1088/0004-637X/701/2/1008}
\bibAnnoteFile{2009ApJ...701.1008C}

\bibitem[{{C{\'o}rsico} et~al.(2019){C{\'o}rsico}, {Althaus}, {Miller
  Bertolami}, and {Kepler}}]{2019A&ARv..27....7C}
{C{\'o}rsico}, A.~H., {Althaus}, L.~G., {Miller Bertolami}, M.~M., and
  {Kepler}, S.~O. (2019).
\newblock {Pulsating white dwarfs: new insights}.
\newblock \emph{\aapr} 27, 7.
\newblock \doi{10.1007/s00159-019-0118-4}
\bibAnnoteFile{2019A&ARv..27....7C}

\bibitem[{{C{\'o}rsico} et~al.(2014){C{\'o}rsico}, {Althaus}, {Miller
  Bertolami}, {Kepler}, and {Garc{\'\i}a-Berro}}]{2014JCAP...08..054C}
{C{\'o}rsico}, A.~H., {Althaus}, L.~G., {Miller Bertolami}, M.~M., {Kepler},
  S.~O., and {Garc{\'\i}a-Berro}, E. (2014).
\newblock {Constraining the neutrino magnetic dipole moment from white dwarf
  pulsations}.
\newblock \emph{\jcap} 2014, 054.
\newblock \doi{10.1088/1475-7516/2014/08/054}
\bibAnnoteFile{2014JCAP...08..054C}

\bibitem[{{C{\'o}rsico} et~al.(2012{\natexlab{b}}){C{\'o}rsico}, {Althaus},
  {Miller Bertolami}, {Romero}, {Garc{\'{\i}}a-Berro}, {Isern}
  et~al.}]{2012MNRAS.424.2792C}
{C{\'o}rsico}, A.~H., {Althaus}, L.~G., {Miller Bertolami}, M.~M., {Romero},
  A.~D., {Garc{\'{\i}}a-Berro}, E., {Isern}, J., et~al. (2012{\natexlab{b}}).
\newblock {The rate of cooling of the pulsating white dwarf star G117-B15A: a
  new asteroseismological inference of the axion mass}.
\newblock \emph{\mnras} 424, 2792--2799.
\newblock \doi{10.1111/j.1365-2966.2012.21401.x}
\bibAnnoteFile{2012MNRAS.424.2792C}

\bibitem[{{C{\'o}rsico} et~al.(2001){C{\'o}rsico}, {Benvenuto}, {Althaus},
  {Isern}, and {Garc{\'{\i}}a-Berro}}]{2001NewA....6..197C}
{C{\'o}rsico}, A.~H., {Benvenuto}, O.~G., {Althaus}, L.~G., {Isern}, J., and
  {Garc{\'{\i}}a-Berro}, E. (2001).
\newblock {The potential of the variable DA white dwarf G117-B15A as a tool for
  fundamental physics}.
\newblock \emph{\nat} 6, 197--213.
\newblock \doi{10.1016/S1384-1076(01)00055-0}
\bibAnnoteFile{2001NewA....6..197C}

\bibitem[{{Cox}(1980)}]{1980tsp..book.....C}
{Cox}, J.~P. (1980).
\newblock \emph{{Theory of stellar pulsation}}
\bibAnnoteFile{1980tsp..book.....C}

\bibitem[{{De Ger{\'o}nimo} et~al.(2019){De Ger{\'o}nimo}, {Battich}, {Miller
  Bertolami}, {Althaus}, and {C{\'o}rsico}}]{2019A&A...630A.100D}
{De Ger{\'o}nimo}, F.~C., {Battich}, T., {Miller Bertolami}, M.~M., {Althaus},
  L.~G., and {C{\'o}rsico}, A.~H. (2019).
\newblock {On the recent parametric determination of an asteroseismological
  model for the DBV star KIC 08626021}.
\newblock \emph{\aap} 630, A100.
\newblock \doi{10.1051/0004-6361/201834988}
\bibAnnoteFile{2019A&A...630A.100D}

\bibitem[{{Dziembowski}(1982)}]{1982AcA....32..147D}
{Dziembowski}, W. (1982).
\newblock {Nonlinear mode coupling in oscillating stars. I - Second order
  theory of the coherent mode coupling}.
\newblock \emph{\actaa} 32, 147--171
\bibAnnoteFile{1982AcA....32..147D}

\bibitem[{{Fields} et~al.(2016){Fields}, {Farmer}, {Petermann}, {Iliadis}, and
  {Timmes}}]{2016ApJ...823...46F}
{Fields}, C.~E., {Farmer}, R., {Petermann}, I., {Iliadis}, C., and {Timmes},
  F.~X. (2016).
\newblock {Properties of Carbon-Oxygen White Dwarfs From Monte Carlo Stellar
  Models}.
\newblock \emph{apj} 823, 46.
\newblock \doi{10.3847/0004-637X/823/1/46}
\bibAnnoteFile{2016ApJ...823...46F}

\bibitem[{{Fontaine} and {Brassard}(2008)}]{2008PASP..120.1043F}
{Fontaine}, G. and {Brassard}, P. (2008).
\newblock {The Pulsating White Dwarf Stars}.
\newblock \emph{PASP} 120, 1043--1096.
\newblock \doi{10.1086/592788}
\bibAnnoteFile{2008PASP..120.1043F}

\bibitem[{{Fontaine} et~al.(2001){Fontaine}, {Brassard}, and
  {Bergeron}}]{2001PASP..113..409F}
{Fontaine}, G., {Brassard}, P., and {Bergeron}, P. (2001).
\newblock {The Potential of White Dwarf Cosmochronology}.
\newblock \emph{\pasp} 113, 409--435.
\newblock \doi{10.1086/319535}
\bibAnnoteFile{2001PASP..113..409F}

\bibitem[{{G{\"a}nsicke} et~al.(2012){G{\"a}nsicke}, {Koester}, {Farihi},
  {Girven}, {Parsons}, and {Breedt}}]{2012MNRAS.424..333G}
{G{\"a}nsicke}, B.~T., {Koester}, D., {Farihi}, J., {Girven}, J., {Parsons},
  S.~G., and {Breedt}, E. (2012).
\newblock {The chemical diversity of exo-terrestrial planetary debris around
  white dwarfs}.
\newblock \emph{\mnras} 424, 333--347.
\newblock \doi{10.1111/j.1365-2966.2012.21201.x}
\bibAnnoteFile{2012MNRAS.424..333G}

\bibitem[{{G{\"a}nsicke} et~al.(2010){G{\"a}nsicke}, {Koester}, {Girven},
  {Marsh}, and {Steeghs}}]{2010Sci...327..188G}
{G{\"a}nsicke}, B.~T., {Koester}, D., {Girven}, J., {Marsh}, T.~R., and
  {Steeghs}, D. (2010).
\newblock {Two White Dwarfs with Oxygen-Rich Atmospheres}.
\newblock \emph{Science} 327, 188.
\newblock \doi{10.1126/science.1180228}
\bibAnnoteFile{2010Sci...327..188G}

\bibitem[{{G{\"a}nsicke} et~al.(2019){G{\"a}nsicke}, {Schreiber}, {Toloza},
  {Fusillo}, {Koester}, and {Manser}}]{2019Natur.576...61G}
{G{\"a}nsicke}, B.~T., {Schreiber}, M.~R., {Toloza}, O., {Fusillo}, N. P.~G.,
  {Koester}, D., and {Manser}, C.~J. (2019).
\newblock {Accretion of a giant planet onto a white dwarf star}.
\newblock \emph{\nat} 576, 61--64.
\newblock \doi{10.1038/s41586-019-1789-8}
\bibAnnoteFile{2019Natur.576...61G}

\bibitem[{{Garc{\'{\i}}a-Berro} and {Oswalt}(2016)}]{2016NewAR..72....1G}
{Garc{\'{\i}}a-Berro}, E. and {Oswalt}, T.~D. (2016).
\newblock {The white dwarf luminosity function}.
\newblock \emph{New Astronomy Reviews} 72, 1--22.
\newblock \doi{10.1016/j.newar.2016.08.001}
\bibAnnoteFile{2016NewAR..72....1G}

\bibitem[{{Garc{\'{\i}}a-Berro} et~al.(2010){Garc{\'{\i}}a-Berro}, {Torres},
  {Althaus}, {Renedo}, {Lor{\'e}n-Aguilar}, {C{\'o}rsico}
  et~al.}]{2010Natur.465..194G}
{Garc{\'{\i}}a-Berro}, E., {Torres}, S., {Althaus}, L.~G., {Renedo}, I.,
  {Lor{\'e}n-Aguilar}, P., {C{\'o}rsico}, A.~H., et~al. (2010).
\newblock {A white dwarf cooling age of 8Gyr for NGC 6791 from physical
  separation processes}.
\newblock \emph{\nat} 465, 194--196.
\newblock \doi{10.1038/nature09045}
\bibAnnoteFile{2010Natur.465..194G}

\bibitem[{{Gentile Fusillo} et~al.(2019){Gentile Fusillo}, {Tremblay},
  {G{\"a}nsicke}, {Manser}, {Cunningham}, {Cukanovaite}
  et~al.}]{2019MNRAS.482.4570G}
{Gentile Fusillo}, N.~P., {Tremblay}, P.-E., {G{\"a}nsicke}, B.~T., {Manser},
  C.~J., {Cunningham}, T., {Cukanovaite}, E., et~al. (2019).
\newblock {A Gaia Data Release 2 catalogue of white dwarfs and a comparison
  with SDSS}.
\newblock \emph{\mnras} 482, 4570--4591.
\newblock \doi{10.1093/mnras/sty3016}
\bibAnnoteFile{2019MNRAS.482.4570G}

\bibitem[{{Giammichele} et~al.(2018){Giammichele}, {Charpinet}, {Fontaine},
  {Brassard}, {Green}, {Van Grootel} et~al.}]{2018Natur.554...73G}
{Giammichele}, N., {Charpinet}, S., {Fontaine}, G., {Brassard}, P., {Green},
  E.~M., {Van Grootel}, V., et~al. (2018).
\newblock {A large oxygen-dominated core from the seismic cartography of a
  pulsating white dwarf}.
\newblock \emph{\nat} 554, 73--76.
\newblock \doi{10.1038/nature25136}
\bibAnnoteFile{2018Natur.554...73G}

\bibitem[{{Gianninas} et~al.(2006){Gianninas}, {Bergeron}, and
  {Fontaine}}]{2006AJ....132..831G}
{Gianninas}, A., {Bergeron}, P., and {Fontaine}, G. (2006).
\newblock {Mapping the ZZ Ceti Instability Strip: Discovery of Six New
  Pulsators}.
\newblock \emph{\aj} 132, 831--835.
\newblock \doi{10.1086/506516}
\bibAnnoteFile{2006AJ....132..831G}

\bibitem[{{Goldreich} and {Wu}(1999)}]{1999ApJ...511..904G}
{Goldreich}, P. and {Wu}, Y. (1999).
\newblock {Gravity Modes in ZZ Ceti Stars. I. Quasi-adiabatic Analysis of
  Overstability}.
\newblock \emph{\apj} 511, 904--915.
\newblock \doi{10.1086/306705}
\bibAnnoteFile{1999ApJ...511..904G}

\bibitem[{{Greiss} et~al.(2015){Greiss}, {G{\"a}nsicke}, {Hermes},
  {Giammichele}, {Fontaine}, {Koester} et~al.}]{2015ASPC..493..169G}
{Greiss}, S., {G{\"a}nsicke}, B.~T., {Hermes}, J.~J., {Giammichele}, N.,
  {Fontaine}, G., {Koester}, D., et~al. (2015).
\newblock \emph{{White Dwarfs in the Kepler Field - What's New?}}, vol. 493 of
  \emph{Astronomical Society of the Pacific Conference Series}.
\newblock 169
\bibAnnoteFile{2015ASPC..493..169G}

\bibitem[{{Greiss} et~al.(2014){Greiss}, {G{\"a}nsicke}, {Hermes}, {Steeghs},
  {Koester}, {Ramsay} et~al.}]{2014MNRAS.438.3086G}
{Greiss}, S., {G{\"a}nsicke}, B.~T., {Hermes}, J.~J., {Steeghs}, D., {Koester},
  D., {Ramsay}, G., et~al. (2014).
\newblock {KIC 11911480: the second ZZ Ceti in the Kepler field}.
\newblock \emph{\mnras} 438, 3086--3092.
\newblock \doi{10.1093/mnras/stt2420}
\bibAnnoteFile{2014MNRAS.438.3086G}

\bibitem[{{Greiss} et~al.(2016){Greiss}, {Hermes}, {G{\"a}nsicke}, {Steeghs},
  {Bell}, {Raddi} et~al.}]{2016MNRAS.457.2855G}
{Greiss}, S., {Hermes}, J.~J., {G{\"a}nsicke}, B.~T., {Steeghs}, D.~T.~H.,
  {Bell}, K.~J., {Raddi}, R., et~al. (2016).
\newblock {The search for ZZ Ceti stars in the original Kepler mission}.
\newblock \emph{\mnras} 457, 2855--2863.
\newblock \doi{10.1093/mnras/stw053}
\bibAnnoteFile{2016MNRAS.457.2855G}

\bibitem[{{Greiss} et~al.(2012){Greiss}, {Steeghs}, {G{\"a}nsicke},
  {Mart{\'\i}n}, {Groot}, {Irwin} et~al.}]{2012AJ....144...24G}
{Greiss}, S., {Steeghs}, D., {G{\"a}nsicke}, B.~T., {Mart{\'\i}n}, E.~L.,
  {Groot}, P.~J., {Irwin}, M.~J., et~al. (2012).
\newblock {Initial Data Release of the Kepler-INT Survey}.
\newblock \emph{\aj} 144, 24.
\newblock \doi{10.1088/0004-6256/144/1/24}
\bibAnnoteFile{2012AJ....144...24G}

\bibitem[{{Hermes} et~al.(2014){Hermes}, {Charpinet}, {Barclay}, {Pakstiene},
  {Mullally}, {Kawaler} et~al.}]{2014ApJ...789...85H}
{Hermes}, J.~J., {Charpinet}, S., {Barclay}, T., {Pakstiene}, E., {Mullally},
  F., {Kawaler}, S.~D., et~al. (2014).
\newblock {Precision Asteroseismology of the Pulsating White Dwarf GD 1212
  Using a Two-wheel-controlled Kepler Spacecraft}.
\newblock \emph{\apj} 789, 85.
\newblock \doi{10.1088/0004-637X/789/1/85}
\bibAnnoteFile{2014ApJ...789...85H}

\bibitem[{{Hermes} et~al.(2015{\natexlab{a}}){Hermes}, {G{\"a}nsicke},
  {Bischoff-Kim}, {Kawaler}, {Fuchs}, {Dunlap} et~al.}]{2015MNRAS.451.1701H}
{Hermes}, J.~J., {G{\"a}nsicke}, B.~T., {Bischoff-Kim}, A., {Kawaler}, S.~D.,
  {Fuchs}, J.~T., {Dunlap}, B.~H., et~al. (2015{\natexlab{a}}).
\newblock {Insights into internal effects of common-envelope evolution using
  the extended Kepler mission}.
\newblock \emph{\mnras} 451, 1701--1712.
\newblock \doi{10.1093/mnras/stv1053}
\bibAnnoteFile{2015MNRAS.451.1701H}

\bibitem[{{Hermes} et~al.(2017{\natexlab{a}}){Hermes}, {G{\"a}nsicke},
  {Kawaler}, {Greiss}, {Tremblay}, {Gentile Fusillo}
  et~al.}]{2017ApJS..232...23H}
{Hermes}, J.~J., {G{\"a}nsicke}, B.~T., {Kawaler}, S.~D., {Greiss}, S.,
  {Tremblay}, P.-E., {Gentile Fusillo}, N.~P., et~al. (2017{\natexlab{a}}).
\newblock {White Dwarf Rotation as a Function of Mass and a Dichotomy of Mode
  Line Widths: Kepler Observations of 27 Pulsating DA White Dwarfs through K2
  Campaign 8}.
\newblock \emph{\apjs} 232, 23.
\newblock \doi{10.3847/1538-4365/aa8bb5}
\bibAnnoteFile{2017ApJS..232...23H}

\bibitem[{{Hermes} et~al.(2017{\natexlab{b}}){Hermes}, {Kawaler},
  {Bischoff-Kim}, {Provencal}, {Dunlap}, and {Clemens}}]{2017ApJ...835..277H}
{Hermes}, J.~J., {Kawaler}, S.~D., {Bischoff-Kim}, A., {Provencal}, J.~L.,
  {Dunlap}, B.~H., and {Clemens}, J.~C. (2017{\natexlab{b}}).
\newblock {A Deep Test of Radial Differential Rotation in a Helium-atmosphere
  White Dwarf. I. Discovery of Pulsations in PG 0112+104}.
\newblock \emph{\apj} 835, 277.
\newblock \doi{10.3847/1538-4357/835/2/277}
\bibAnnoteFile{2017ApJ...835..277H}

\bibitem[{{Hermes} et~al.(2017{\natexlab{c}}){Hermes}, {Kawaler}, {Romero},
  {Kepler}, {Tremblay}, {Bell} et~al.}]{2017ApJ...841L...2H}
{Hermes}, J.~J., {Kawaler}, S.~D., {Romero}, A.~D., {Kepler}, S.~O.,
  {Tremblay}, P.-E., {Bell}, K.~J., et~al. (2017{\natexlab{c}}).
\newblock {Evidence from K2 for Rapid Rotation in the Descendant of an
  Intermediate-mass Star}.
\newblock \emph{\apjl} 841, L2.
\newblock \doi{10.3847/2041-8213/aa6ffc}
\bibAnnoteFile{2017ApJ...841L...2H}

\bibitem[{{Hermes} et~al.(2015{\natexlab{b}}){Hermes}, {Montgomery}, {Bell},
  {Chote}, {G{\"a}nsicke}, {Kawaler} et~al.}]{2015ApJ...810L...5H}
{Hermes}, J.~J., {Montgomery}, M.~H., {Bell}, K.~J., {Chote}, P.,
  {G{\"a}nsicke}, B.~T., {Kawaler}, S.~D., et~al. (2015{\natexlab{b}}).
\newblock {A Second Case of Outbursts in a Pulsating White Dwarf Observed by
  Kepler}.
\newblock \emph{\apjl} 810, L5.
\newblock \doi{10.1088/2041-8205/810/1/L5}
\bibAnnoteFile{2015ApJ...810L...5H}

\bibitem[{{Hermes} et~al.(2011){Hermes}, {Mullally}, {{\O}stensen}, {Williams},
  {Telting}, {Southworth} et~al.}]{2011ApJ...741L..16H}
{Hermes}, J.~J., {Mullally}, F., {{\O}stensen}, R.~H., {Williams}, K.~A.,
  {Telting}, J., {Southworth}, J., et~al. (2011).
\newblock {Discovery of a ZZ Ceti in the Kepler Mission Field}.
\newblock \emph{\apjl} 741, L16.
\newblock \doi{10.1088/2041-8205/741/1/L16}
\bibAnnoteFile{2011ApJ...741L..16H}

\bibitem[{{Hollands} et~al.(2018){Hollands}, {G{\"a}nsicke}, and
  {Koester}}]{2018MNRAS.477...93H}
{Hollands}, M.~A., {G{\"a}nsicke}, B.~T., and {Koester}, D. (2018).
\newblock {Cool DZ white dwarfs II: compositions and evolution of old remnant
  planetary systems}.
\newblock \emph{\mnras} 477, 93--111.
\newblock \doi{10.1093/mnras/sty592}
\bibAnnoteFile{2018MNRAS.477...93H}

\bibitem[{{Hollands} et~al.(2020){Hollands}, {Tremblay}, {G{\"a}nsicke},
  {Camisassa}, {Koester}, {Aungwerojwit} et~al.}]{2020NatAs.tmp....3H}
{Hollands}, M.~A., {Tremblay}, P.~E., {G{\"a}nsicke}, B.~T., {Camisassa},
  M.~E., {Koester}, D., {Aungwerojwit}, A., et~al. (2020).
\newblock {An ultra-massive white dwarf with a mixed hydrogen-carbon atmosphere
  as a likely merger remnant}.
\newblock \emph{Nature Astronomy} \doi{10.1038/s41550-020-1028-0}
\bibAnnoteFile{2020NatAs.tmp....3H}

\bibitem[{{Howell} et~al.(2014){Howell}, {Sobeck}, {Haas}, {Still}, {Barclay},
  {Mullally} et~al.}]{2014PASP..126..398H}
{Howell}, S.~B., {Sobeck}, C., {Haas}, M., {Still}, M., {Barclay}, T.,
  {Mullally}, F., et~al. (2014).
\newblock {The K2 Mission: Characterization and Early Results}.
\newblock \emph{PASP} 126, 398.
\newblock \doi{10.1086/676406}
\bibAnnoteFile{2014PASP..126..398H}

\bibitem[{{Isern} et~al.(2008){Isern}, {Garc{\'\i}a-Berro}, {Torres}, and
  {Catal{\'a}n}}]{2008ApJ...682L.109I}
{Isern}, J., {Garc{\'\i}a-Berro}, E., {Torres}, S., and {Catal{\'a}n}, S.
  (2008).
\newblock {Axions and the Cooling of White Dwarf Stars}.
\newblock \emph{\apjl} 682, L109.
\newblock \doi{10.1086/591042}
\bibAnnoteFile{2008ApJ...682L.109I}

\bibitem[{{Isern} et~al.(2018){Isern}, {Garc{\'\i}a-Berro}, {Torres},
  {Cojocaru}, and {Catal{\'a}n}}]{2018MNRAS.478.2569I}
{Isern}, J., {Garc{\'\i}a-Berro}, E., {Torres}, S., {Cojocaru}, R., and
  {Catal{\'a}n}, S. (2018).
\newblock {Axions and the luminosity function of white dwarfs: the thin and
  thick discs, and the halo}.
\newblock \emph{\mnras} 478, 2569--2575.
\newblock \doi{10.1093/mnras/sty1162}
\bibAnnoteFile{2018MNRAS.478.2569I}

\bibitem[{{Isern} et~al.(1992){Isern}, {Hernanz}, and
  {Garcia-Berro}}]{1992ApJ...392L..23I}
{Isern}, J., {Hernanz}, M., and {Garcia-Berro}, E. (1992).
\newblock {Axion cooling of white dwarfs}.
\newblock \emph{ApJl} 392, L23--L25.
\newblock \doi{10.1086/186416}
\bibAnnoteFile{1992ApJ...392L..23I}

\bibitem[{{Kawaler}(2004)}]{2004IAUS..215..561K}
{Kawaler}, S.~D. (2004).
\newblock {White Dwarf Rotation: Observations and Theory (Invited Review)}.
\newblock In \emph{Stellar Rotation}, eds. A.~{Maeder} and P.~{Eenens}. vol.
  215 of \emph{IAU Symposium}, 561
\bibAnnoteFile{2004IAUS..215..561K}

\bibitem[{{Kawaler} et~al.(1985){Kawaler}, {Winget}, and
  {Hansen}}]{1985ApJ...295..547K}
{Kawaler}, S.~D., {Winget}, D.~E., and {Hansen}, C.~J. (1985).
\newblock {Evolution of the pulsation properties of hot pre-white dwarf stars.}
\newblock \emph{\apj} 295, 547--560.
\newblock \doi{10.1086/163398}
\bibAnnoteFile{1985ApJ...295..547K}

\bibitem[{{Kepler}(1984)}]{1984ApJ...286..314K}
{Kepler}, S.~O. (1984).
\newblock {Light and line profile variations die to r-mode pusations with an
  application to the ZZ Ceti star G 117-B15A.}
\newblock \emph{\apj} 286, 314--327.
\newblock \doi{10.1086/162601}
\bibAnnoteFile{1984ApJ...286..314K}

\bibitem[{{Kepler} et~al.(2016{\natexlab{a}}){Kepler}, {Koester}, and
  {Ourique}}]{2016Sci...352...67K}
{Kepler}, S.~O., {Koester}, D., and {Ourique}, G. (2016{\natexlab{a}}).
\newblock {A white dwarf with an oxygen atmosphere}.
\newblock \emph{Science} 352, 67--69.
\newblock \doi{10.1126/science.aad6705}
\bibAnnoteFile{2016Sci...352...67K}

\bibitem[{{Kepler} et~al.(2016{\natexlab{b}}){Kepler}, {Pelisoli}, {Koester},
  {Ourique}, {Romero}, {Reindl} et~al.}]{2016MNRAS.455.3413K}
{Kepler}, S.~O., {Pelisoli}, I., {Koester}, D., {Ourique}, G., {Romero}, A.~D.,
  {Reindl}, N., et~al. (2016{\natexlab{b}}).
\newblock {New white dwarf and subdwarf stars in the Sloan Digital Sky Survey
  Data Release 12}.
\newblock \emph{\mnras} 455, 3413--3423.
\newblock \doi{10.1093/mnras/stv2526}
\bibAnnoteFile{2016MNRAS.455.3413K}

\bibitem[{{Kepler} et~al.(2019){Kepler}, {Pelisoli}, {Koester}, {Reindl},
  {Geier}, {Romero} et~al.}]{2019MNRAS.486.2169K}
{Kepler}, S.~O., {Pelisoli}, I., {Koester}, D., {Reindl}, N., {Geier}, S.,
  {Romero}, A.~D., et~al. (2019).
\newblock {White dwarf and subdwarf stars in the Sloan Digital Sky Survey Data
  Release 14}.
\newblock \emph{\mnras} 486, 2169--2183.
\newblock \doi{10.1093/mnras/stz960}
\bibAnnoteFile{2019MNRAS.486.2169K}

\bibitem[{{Kilic} et~al.(2017){Kilic}, {Munn}, {Harris}, {von Hippel},
  {Liebert}, {Williams} et~al.}]{2017ApJ...837..162K}
{Kilic}, M., {Munn}, J.~A., {Harris}, H.~C., {von Hippel}, T., {Liebert},
  J.~W., {Williams}, K.~A., et~al. (2017).
\newblock {The Ages of the Thin Disk, Thick Disk, and the Halo from Nearby
  White Dwarfs}.
\newblock \emph{apj} 837, 162.
\newblock \doi{10.3847/1538-4357/aa62a5}
\bibAnnoteFile{2017ApJ...837..162K}

\bibitem[{{Kleinman} et~al.(2013){Kleinman}, {Kepler}, {Koester}, {Pelisoli},
  {Pe{\c c}anha}, {Nitta} et~al.}]{2013ApJS..204....5K}
{Kleinman}, S.~J., {Kepler}, S.~O., {Koester}, D., {Pelisoli}, I., {Pe{\c
  c}anha}, V., {Nitta}, A., et~al. (2013).
\newblock {SDSS DR7 White Dwarf Catalog}.
\newblock \emph{ApJs} 204, 5.
\newblock \doi{10.1088/0067-0049/204/1/5}
\bibAnnoteFile{2013ApJS..204....5K}

\bibitem[{{Koester}(2010)}]{2010MmSAI..81..921K}
{Koester}, D. (2010).
\newblock {White dwarf spectra and atmosphere models}.
\newblock \emph{\memsai} 81, 921--931
\bibAnnoteFile{2010MmSAI..81..921K}

\bibitem[{{Koester} et~al.(2014){Koester}, {G{\"a}nsicke}, and
  {Farihi}}]{2014A&A...566A..34K}
{Koester}, D., {G{\"a}nsicke}, B.~T., and {Farihi}, J. (2014).
\newblock {The frequency of planetary debris around young white dwarfs}.
\newblock \emph{\aap} 566, A34.
\newblock \doi{10.1051/0004-6361/201423691}
\bibAnnoteFile{2014A&A...566A..34K}

\bibitem[{{Kunz} et~al.(2002){Kunz}, {Fey}, {Jaeger}, {Mayer}, {Hammer},
  {Staudt} et~al.}]{2002ApJ...567..643K}
{Kunz}, R., {Fey}, M., {Jaeger}, M., {Mayer}, A., {Hammer}, J.~W., {Staudt},
  G., et~al. (2002).
\newblock {Astrophysical Reaction Rate of $^{12}$C({$\alpha$},
  {$\gamma$})$^{16}$O}.
\newblock \emph{\apj} 567, 643--650.
\newblock \doi{10.1086/338384}
\bibAnnoteFile{2002ApJ...567..643K}

\bibitem[{{Kurtz} et~al.(2008){Kurtz}, {Shibahashi}, {Dhillon}, {Marsh}, and
  {Littlefair}}]{2008MNRAS.389.1771K}
{Kurtz}, D.~W., {Shibahashi}, H., {Dhillon}, V.~S., {Marsh}, T.~R., and
  {Littlefair}, S.~P. (2008).
\newblock {A search for a new class of pulsating DA white dwarf stars in the DB
  gap}.
\newblock \emph{\mnras} 389, 1771--1779.
\newblock \doi{10.1111/j.1365-2966.2008.13664.x}
\bibAnnoteFile{2008MNRAS.389.1771K}

\bibitem[{{Kurtz} et~al.(2013){Kurtz}, {Shibahashi}, {Dhillon}, {Marsh},
  {Littlefair}, {Copperwheat} et~al.}]{2013MNRAS.432.1632K}
{Kurtz}, D.~W., {Shibahashi}, H., {Dhillon}, V.~S., {Marsh}, T.~R.,
  {Littlefair}, S.~P., {Copperwheat}, C.~M., et~al. (2013).
\newblock {Hot DAVs: a probable new class of pulsating white dwarf stars}.
\newblock \emph{\mnras} 432, 1632--1639.
\newblock \doi{10.1093/mnras/stt585}
\bibAnnoteFile{2013MNRAS.432.1632K}

\bibitem[{{Landolt}(1968)}]{1968ApJ...153..151L}
{Landolt}, A.~U. (1968).
\newblock {A New Short-Period Blue Variable}.
\newblock \emph{\apj} 153, 151.
\newblock \doi{10.1086/149645}
\bibAnnoteFile{1968ApJ...153..151L}

\bibitem[{{Luan} and {Goldreich}(2018)}]{2018ApJ...863...82L}
{Luan}, J. and {Goldreich}, P. (2018).
\newblock {DAVs: Red Edge and Outbursts}.
\newblock \emph{\apj} 863, 82.
\newblock \doi{10.3847/1538-4357/aad0f4}
\bibAnnoteFile{2018ApJ...863...82L}

\bibitem[{{Maeda} and {Shibahashi}(2014)}]{2014PASJ...66...76M}
{Maeda}, K. and {Shibahashi}, H. (2014).
\newblock {Pulsations of pre-white dwarfs with hydrogen-dominated atmospheres}.
\newblock \emph{\pasj} 66, 76.
\newblock \doi{10.1093/pasj/psu051}
\bibAnnoteFile{2014PASJ...66...76M}

\bibitem[{{Maoz} et~al.(2014){Maoz}, {Mannucci}, and
  {Nelemans}}]{2014ARA&A..52..107M}
{Maoz}, D., {Mannucci}, F., and {Nelemans}, G. (2014).
\newblock {Observational Clues to the Progenitors of Type Ia Supernovae}.
\newblock \emph{Annula Review Astronomy and Astrophysics} 52, 107--170.
\newblock \doi{10.1146/annurev-astro-082812-141031}
\bibAnnoteFile{2014ARA&A..52..107M}

\bibitem[{{McGraw}(1979)}]{1979ApJ...229..203M}
{McGraw}, J.~T. (1979).
\newblock {The physical properties of the ZZ Ceti stars and their pulsations.}
\newblock \emph{\apj} 229, 203--211.
\newblock \doi{10.1086/156946}
\bibAnnoteFile{1979ApJ...229..203M}

\bibitem[{{Mestel}(1952)}]{1952MNRAS.112..583M}
{Mestel}, L. (1952).
\newblock {On the theory of white dwarf stars. I. The energy sources of white
  dwarfs}.
\newblock \emph{\mnras} 112, 583.
\newblock \doi{10.1093/mnras/112.6.583}
\bibAnnoteFile{1952MNRAS.112..583M}

\bibitem[{{Metcalfe}(2003)}]{2003ApJ...587L..43M}
{Metcalfe}, T.~S. (2003).
\newblock {White Dwarf Asteroseismology and the
  $^{12}$C({\ensuremath{\alpha}},{\ensuremath{\gamma}})$^{16}$O Rate}.
\newblock \emph{\apjl} 587, L43--L46.
\newblock \doi{10.1086/375044}
\bibAnnoteFile{2003ApJ...587L..43M}

\bibitem[{{Miller Bertolami}(2014)}]{2014A&A...562A.123M}
{Miller Bertolami}, M.~M. (2014).
\newblock {Limits on the neutrino magnetic dipole moment from the luminosity
  function of hot white dwarfs}.
\newblock \emph{\aap} 562, A123.
\newblock \doi{10.1051/0004-6361/201322641}
\bibAnnoteFile{2014A&A...562A.123M}

\bibitem[{{Miller Bertolami} and {Althaus}(2006)}]{2006A&A...454..845M}
{Miller Bertolami}, M.~M. and {Althaus}, L.~G. (2006).
\newblock {Full evolutionary models for PG 1159 stars. Implications for the
  helium-rich O(He) stars}.
\newblock \emph{\aap} 454, 845--854.
\newblock \doi{10.1051/0004-6361:20054723}
\bibAnnoteFile{2006A&A...454..845M}

\bibitem[{{Miller Bertolami} et~al.(2014){Miller Bertolami}, {Melendez},
  {Althaus}, and {Isern}}]{2014JCAP...10..069M}
{Miller Bertolami}, M.~M., {Melendez}, B.~E., {Althaus}, L.~G., and {Isern}, J.
  (2014).
\newblock {Revisiting the axion bounds from the Galactic white dwarf luminosity
  function}.
\newblock \emph{\jcap} 2014, 069.
\newblock \doi{10.1088/1475-7516/2014/10/069}
\bibAnnoteFile{2014JCAP...10..069M}

\bibitem[{{Montgomery} et~al.(2020){Montgomery}, {Hermes}, {Winget}, {Dunlap},
  and {Bell}}]{2020ApJ...890...11M}
{Montgomery}, M.~H., {Hermes}, J.~J., {Winget}, D.~E., {Dunlap}, B.~H., and
  {Bell}, K.~J. (2020).
\newblock {Limits on Mode Coherence in Pulsating DA White Dwarfs Due to a
  Nonstatic Convection Zone}.
\newblock \emph{\apj} 890, 11.
\newblock \doi{10.3847/1538-4357/ab6a0e}
\bibAnnoteFile{2020ApJ...890...11M}

\bibitem[{{Montgomery} et~al.(1999){Montgomery}, {Klumpe}, {Winget}, and
  {Wood}}]{1999ApJ...525..482M}
{Montgomery}, M.~H., {Klumpe}, E.~W., {Winget}, D.~E., and {Wood}, M.~A.
  (1999).
\newblock {Evolutionary Calculations of Phase Separation in Crystallizing White
  Dwarf Stars}.
\newblock \emph{\apj} 525, 482--491.
\newblock \doi{10.1086/307871}
\bibAnnoteFile{1999ApJ...525..482M}

\bibitem[{{Montgomery} et~al.(2008){Montgomery}, {Williams}, {Winget},
  {Dufour}, {DeGennaro}, and {Liebert}}]{2008ApJ...678L..51M}
{Montgomery}, M.~H., {Williams}, K.~A., {Winget}, D.~E., {Dufour}, P.,
  {DeGennaro}, S., and {Liebert}, J. (2008).
\newblock {SDSS J142625.71+575218.3: A Prototype for a New Class of Variable
  White Dwarf}.
\newblock \emph{\apjl} 678, L51.
\newblock \doi{10.1086/588286}
\bibAnnoteFile{2008ApJ...678L..51M}

\bibitem[{{Montgomery} and {Winget}(1999)}]{1999ApJ...526..976M}
{Montgomery}, M.~H. and {Winget}, D.~E. (1999).
\newblock {The Effect of Crystallization on the Pulsations of White Dwarf
  Stars}.
\newblock \emph{\apj} 526, 976--990.
\newblock \doi{10.1086/308044}
\bibAnnoteFile{1999ApJ...526..976M}

\bibitem[{{Moya} et~al.(2018){Moya}, {Barcel{\'o} Forteza}, {Bonfanti},
  {Salmon}, {Van Grootel}, and {Barrado}}]{2018A&A...620A.203M}
{Moya}, A., {Barcel{\'o} Forteza}, S., {Bonfanti}, A., {Salmon}, S.~J.~A.~J.,
  {Van Grootel}, V., and {Barrado}, D. (2018).
\newblock {Asteroseismic potential of CHEOPS}.
\newblock \emph{\aap} 620, A203.
\newblock \doi{10.1051/0004-6361/201833772}
\bibAnnoteFile{2018A&A...620A.203M}

\bibitem[{{Nather} et~al.(1990){Nather}, {Winget}, {Clemens}, {Hansen}, and
  {Hine}}]{1990ApJ...361..309N}
{Nather}, R.~E., {Winget}, D.~E., {Clemens}, J.~C., {Hansen}, C.~J., and
  {Hine}, B.~P. (1990).
\newblock {The whole earth telescope - A new astronomical instrument}.
\newblock \emph{\apj} 361, 309--317.
\newblock \doi{10.1086/169196}
\bibAnnoteFile{1990ApJ...361..309N}

\bibitem[{{{\O}stensen} et~al.(2011{\natexlab{a}}){{\O}stensen}, {Bloemen},
  {Vu{\v c}kovi{\'c}}, {Aerts}, {Oreiro}, {Kinemuchi}
  et~al.}]{2011ApJ...736L..39O}
{{\O}stensen}, R.~H., {Bloemen}, S., {Vu{\v c}kovi{\'c}}, M., {Aerts}, C.,
  {Oreiro}, R., {Kinemuchi}, K., et~al. (2011{\natexlab{a}}).
\newblock {At Last A V777 Her Pulsator in the Kepler Field}.
\newblock \emph{\apjl} 736, L39.
\newblock \doi{10.1088/2041-8205/736/2/L39}
\bibAnnoteFile{2011ApJ...736L..39O}

\bibitem[{{{\O}stensen} et~al.(2011{\natexlab{b}}){{\O}stensen}, {Silvotti},
  {Charpinet}, {Oreiro}, {Bloemen}, {Baran} et~al.}]{2011MNRAS.414.2860O}
{{\O}stensen}, R.~H., {Silvotti}, R., {Charpinet}, S., {Oreiro}, R., {Bloemen},
  S., {Baran}, A.~S., et~al. (2011{\natexlab{b}}).
\newblock {First Kepler results on compact pulsators - VI. Targets in the final
  half of the survey phase}.
\newblock \emph{\mnras} 414, 2860--2870.
\newblock \doi{10.1111/j.1365-2966.2011.18405.x}
\bibAnnoteFile{2011MNRAS.414.2860O}

\bibitem[{{Papaloizou} and {Pringle}(1978)}]{1978MNRAS.182..423P}
{Papaloizou}, J. and {Pringle}, J.~E. (1978).
\newblock {Non-radial oscillations of rotating stars and their relevance to the
  short-period oscillations of cataclysmic variables.}
\newblock \emph{\mnras} 182, 423--442.
\newblock \doi{10.1093/mnras/182.3.423}
\bibAnnoteFile{1978MNRAS.182..423P}

\bibitem[{{Piotto}(2018)}]{2018EPSC...12..969P}
{Piotto}, G. (2018).
\newblock {PLATO Science: main goals and expected achievements}.
\newblock In \emph{European Planetary Science Congress}. EPSC2018--969
\bibAnnoteFile{2018EPSC...12..969P}

\bibitem[{{Pyrzas} et~al.(2015){Pyrzas}, {G{\"a}nsicke}, {Hermes},
  {Copperwheat}, {Rebassa-Mansergas}, {Dhillon} et~al.}]{2015MNRAS.447..691P}
{Pyrzas}, S., {G{\"a}nsicke}, B.~T., {Hermes}, J.~J., {Copperwheat}, C.~M.,
  {Rebassa-Mansergas}, A., {Dhillon}, V.~S., et~al. (2015).
\newblock {Discovery of ZZ Cetis in detached white dwarf plus main-sequence
  binaries}.
\newblock \emph{\mnras} 447, 691--697.
\newblock \doi{10.1093/mnras/stu2412}
\bibAnnoteFile{2015MNRAS.447..691P}

\bibitem[{{Reding} et~al.(2020){Reding}, {Hermes}, {Vanderbosch}, {Dennihy},
  {Kaiser}, {Mace} et~al.}]{2020ApJ...894...19R}
{Reding}, J.~S., {Hermes}, J.~J., {Vanderbosch}, Z., {Dennihy}, E., {Kaiser},
  B.~C., {Mace}, C.~B., et~al. (2020).
\newblock {An Isolated White Dwarf with 317 s Rotation and Magnetic Emission}.
\newblock \emph{\apj} 894, 19.
\newblock \doi{10.3847/1538-4357/ab8239}
\bibAnnoteFile{2020ApJ...894...19R}

\bibitem[{{Ricker} et~al.(2015){Ricker}, {Winn}, {Vanderspek}, {Latham},
  {Bakos}, {Bean} et~al.}]{2015JATIS...1a4003R}
{Ricker}, G.~R., {Winn}, J.~N., {Vanderspek}, R., {Latham}, D.~W., {Bakos},
  G.~{\'A}., {Bean}, J.~L., et~al. (2015).
\newblock {Transiting Exoplanet Survey Satellite (TESS)}.
\newblock \emph{Journal of Astronomical Telescopes, Instruments, and Systems}
  1, 014003.
\newblock \doi{10.1117/1.JATIS.1.1.014003}
\bibAnnoteFile{2015JATIS...1a4003R}

\bibitem[{{Robinson} et~al.(1982){Robinson}, {Kepler}, and
  {Nather}}]{1982ApJ...259..219R}
{Robinson}, E.~L., {Kepler}, S.~O., and {Nather}, R.~E. (1982).
\newblock {Multicolor variations of the ZZ Ceti stars}.
\newblock \emph{\apj} 259, 219--231.
\newblock \doi{10.1086/160162}
\bibAnnoteFile{1982ApJ...259..219R}

\bibitem[{{Romero} et~al.(2017){Romero}, {C{\'o}rsico}, {Castanheira}, {De
  Ger{\'o}nimo}, {Kepler}, {Koester} et~al.}]{2017ApJ...851...60R}
{Romero}, A.~D., {C{\'o}rsico}, A.~H., {Castanheira}, B.~G., {De Ger{\'o}nimo},
  F.~C., {Kepler}, S.~O., {Koester}, D., et~al. (2017).
\newblock {Probing the Structure of Kepler ZZ Ceti Stars with Full Evolutionary
  Models-based Asteroseismology}.
\newblock \emph{\apj} 851, 60.
\newblock \doi{10.3847/1538-4357/aa9899}
\bibAnnoteFile{2017ApJ...851...60R}

\bibitem[{{Saio}(2013)}]{2013EPJWC..4305005S}
{Saio}, H. (2013).
\newblock {Pulsations in white dwarfs: Selected topics}.
\newblock In \emph{European Physical Journal Web of Conferences}. vol.~43 of
  \emph{European Physical Journal Web of Conferences}, 05005.
\newblock \doi{10.1051/epjconf/20134305005}
\bibAnnoteFile{2013EPJWC..4305005S}

\bibitem[{{Salaris} et~al.(2009){Salaris}, {Serenelli}, {Weiss}, and {Miller
  Bertolami}}]{2009ApJ...692.1013S}
{Salaris}, M., {Serenelli}, A., {Weiss}, A., and {Miller Bertolami}, M. (2009).
\newblock {Semi-empirical White Dwarf Initial-Final Mass Relationships: A
  Thorough Analysis of Systematic Uncertainties Due to Stellar Evolution
  Models}.
\newblock \emph{\apj} 692, 1013--1032.
\newblock \doi{10.1088/0004-637X/692/2/1013}
\bibAnnoteFile{2009ApJ...692.1013S}

\bibitem[{{Shibahashi}(2005)}]{2005EAS....17..143S}
{Shibahashi}, H. (2005).
\newblock {The DB gap and pulsations of white dwarfs}.
\newblock In \emph{EAS Publications Series}, eds. G.~{Alecian}, O.~{Richard},
  and S.~{Vauclair}. vol.~17 of \emph{EAS Publications Series}, 143--148.
\newblock \doi{10.1051/eas:2005108}
\bibAnnoteFile{2005EAS....17..143S}

\bibitem[{{Shibahashi}(2007)}]{2007AIPC..948...35S}
{Shibahashi}, H. (2007).
\newblock {The DB Gap and Pulsations of White Dwarfs}.
\newblock In \emph{Unsolved Problems in Stellar Physics: A Conference in Honor
  of Douglas Gough}, eds. R.~J. {Stancliffe}, G.~{Houdek}, R.~G. {Martin}, and
  C.~A. {Tout}. vol. 948 of \emph{American Institute of Physics Conference
  Series}, 35--42.
\newblock \doi{10.1063/1.2818994}
\bibAnnoteFile{2007AIPC..948...35S}

\bibitem[{{Starrfield} et~al.(1984){Starrfield}, {Cox}, {Kidman}, and
  {Pesnell}}]{1984ApJ...281..800S}
{Starrfield}, S., {Cox}, A.~N., {Kidman}, R.~B., and {Pesnell}, W.~D. (1984).
\newblock {Nonradial instability strips based on carbon and oxygen partial
  ionization in hot, evolved stars.}
\newblock \emph{\apj} 281, 800--810.
\newblock \doi{10.1086/162158}
\bibAnnoteFile{1984ApJ...281..800S}

\bibitem[{{Straniero} et~al.(2003){Straniero}, {Dom{\'\i}nguez}, {Imbriani},
  and {Piersanti}}]{2003ApJ...583..878S}
{Straniero}, O., {Dom{\'\i}nguez}, I., {Imbriani}, G., and {Piersanti}, L.
  (2003).
\newblock {The Chemical Composition of White Dwarfs as a Test of Convective
  Efficiency during Core Helium Burning}.
\newblock \emph{\apj} 583, 878--884.
\newblock \doi{10.1086/345427}
\bibAnnoteFile{2003ApJ...583..878S}

\bibitem[{{Thompson} et~al.(2018){Thompson}, {Coughlin}, {Hoffman}, {Mullally},
  {Christiansen}, {Burke} et~al.}]{2018ApJS..235...38T}
{Thompson}, S.~E., {Coughlin}, J.~L., {Hoffman}, K., {Mullally}, F.,
  {Christiansen}, J.~L., {Burke}, C.~J., et~al. (2018).
\newblock {Planetary Candidates Observed by Kepler. VIII. A Fully Automated
  Catalog with Measured Completeness and Reliability Based on Data Release 25}.
\newblock \emph{\apjs} 235, 38.
\newblock \doi{10.3847/1538-4365/aab4f9}
\bibAnnoteFile{2018ApJS..235...38T}

\bibitem[{{Timmes} et~al.(2018){Timmes}, {Townsend}, {Bauer}, {Thoul},
  {Fields}, and {Wolf}}]{2018ApJ...867L..30T}
{Timmes}, F.~X., {Townsend}, R. H.~D., {Bauer}, E.~B., {Thoul}, A., {Fields},
  C.~E., and {Wolf}, W.~M. (2018).
\newblock {The Impact of White Dwarf Luminosity Profiles on Oscillation
  Frequencies}.
\newblock \emph{\apjl} 867, L30.
\newblock \doi{10.3847/2041-8213/aae70f}
\bibAnnoteFile{2018ApJ...867L..30T}

\bibitem[{{Tremblay} et~al.(2016){Tremblay}, {Cummings}, {Kalirai},
  {G{\"a}nsicke}, {Gentile-Fusillo}, and {Raddi}}]{2016MNRAS.461.2100T}
{Tremblay}, P.~E., {Cummings}, J., {Kalirai}, J.~S., {G{\"a}nsicke}, B.~T.,
  {Gentile-Fusillo}, N., and {Raddi}, R. (2016).
\newblock {The field white dwarf mass distribution}.
\newblock \emph{\mnras} 461, 2100--2114.
\newblock \doi{10.1093/mnras/stw1447}
\bibAnnoteFile{2016MNRAS.461.2100T}

\bibitem[{{Tremblay} et~al.(2019){Tremblay}, {Fontaine}, {Fusillo}, {Dunlap},
  {G{\"a}nsicke}, {Hollands} et~al.}]{2019Natur.565..202T}
{Tremblay}, P.-E., {Fontaine}, G., {Fusillo}, N.~P.~G., {Dunlap}, B.~H.,
  {G{\"a}nsicke}, B.~T., {Hollands}, M.~A., et~al. (2019).
\newblock {Core crystallization and pile-up in the cooling sequence of evolving
  white dwarfs}.
\newblock \emph{\nat} 565, 202--205.
\newblock \doi{10.1038/s41586-018-0791-x}
\bibAnnoteFile{2019Natur.565..202T}

\bibitem[{{Unno} et~al.(1989){Unno}, {Osaki}, {Ando}, {Saio}, and
  {Shibahashi}}]{1989nos..book.....U}
{Unno}, W., {Osaki}, Y., {Ando}, H., {Saio}, H., and {Shibahashi}, H. (1989).
\newblock \emph{{Nonradial oscillations of stars}}
\bibAnnoteFile{1989nos..book.....U}

\bibitem[{{Van Grootel} et~al.(2012){Van Grootel}, {Dupret}, {Fontaine},
  {Brassard}, {Grigahc{\`e}ne}, and {Quirion}}]{2012A&A...539A..87V}
{Van Grootel}, V., {Dupret}, M.-A., {Fontaine}, G., {Brassard}, P.,
  {Grigahc{\`e}ne}, A., and {Quirion}, P.-O. (2012).
\newblock {The instability strip of ZZ Ceti white dwarfs. I. Introduction of
  time-dependent convection}.
\newblock \emph{\aap} 539, A87.
\newblock \doi{10.1051/0004-6361/201118371}
\bibAnnoteFile{2012A&A...539A..87V}

\bibitem[{{Van Horn}(2015)}]{2015uswd.book.....V}
{Van Horn}, H.~M. (2015).
\newblock \emph{{Unlocking the Secrets of White Dwarf Stars}}.
\newblock \doi{10.1007/978-3-319-09369-7}
\bibAnnoteFile{2015uswd.book.....V}

\bibitem[{{Winget} and {Kepler}(2008)}]{2008ARA&A..46..157W}
{Winget}, D.~E. and {Kepler}, S.~O. (2008).
\newblock {Pulsating White Dwarf Stars and Precision Asteroseismology}.
\newblock \emph{\araa} 46, 157--199.
\newblock \doi{10.1146/annurev.astro.46.060407.145250}
\bibAnnoteFile{2008ARA&A..46..157W}

\bibitem[{{Winget} et~al.(2009){Winget}, {Kepler}, {Campos}, {Montgomery},
  {Girardi}, {Bergeron} et~al.}]{2009ApJ...693L...6W}
{Winget}, D.~E., {Kepler}, S.~O., {Campos}, F., {Montgomery}, M.~H., {Girardi},
  L., {Bergeron}, P., et~al. (2009).
\newblock {The Physics of Crystallization From Globular Cluster White Dwarf
  Stars in NGC 6397}.
\newblock \emph{\apjl} 693, L6--L10.
\newblock \doi{10.1088/0004-637X/693/1/L6}
\bibAnnoteFile{2009ApJ...693L...6W}

\bibitem[{{Winget} et~al.(1991){Winget}, {Nather}, {Clemens}, {Provencal},
  {Kleinman}, {Bradley} et~al.}]{1991ApJ...378..326W}
{Winget}, D.~E., {Nather}, R.~E., {Clemens}, J.~C., {Provencal}, J.,
  {Kleinman}, S.~J., {Bradley}, P.~A., et~al. (1991).
\newblock {Asteroseismology of the DOV star PG 1159 - 035 with the Whole Earth
  Telescope}.
\newblock \emph{\apj} 378, 326--346.
\newblock \doi{10.1086/170434}
\bibAnnoteFile{1991ApJ...378..326W}

\bibitem[{{Winget} et~al.(1994){Winget}, {Nather}, {Clemens}, {Provencal},
  {Kleinman}, {Bradley} et~al.}]{1994ApJ...430..839W}
{Winget}, D.~E., {Nather}, R.~E., {Clemens}, J.~C., {Provencal}, J.~L.,
  {Kleinman}, S.~J., {Bradley}, P.~A., et~al. (1994).
\newblock {Whole earth telescope observations of the DBV white dwarf GD 358}.
\newblock \emph{\apj} 430, 839--849.
\newblock \doi{10.1086/174455}
\bibAnnoteFile{1994ApJ...430..839W}

\bibitem[{{Winget} et~al.(2004){Winget}, {Sullivan}, {Metcalfe}, {Kawaler}, and
  {Montgomery}}]{2004ApJ...602L.109W}
{Winget}, D.~E., {Sullivan}, D.~J., {Metcalfe}, T.~S., {Kawaler}, S.~D., and
  {Montgomery}, M.~H. (2004).
\newblock {A Strong Test of Electroweak Theory Using Pulsating DB White Dwarf
  Stars as Plasmon Neutrino Detectors}.
\newblock \emph{\apjl} 602, L109--L112.
\newblock \doi{10.1086/382591}
\bibAnnoteFile{2004ApJ...602L.109W}

\bibitem[{{Winget} et~al.(1982){Winget}, {van Horn}, {Tassoul}, {Fontaine},
  {Hansen}, and {Carroll}}]{1982ApJ...252L..65W}
{Winget}, D.~E., {van Horn}, H.~M., {Tassoul}, M., {Fontaine}, G., {Hansen},
  C.~J., and {Carroll}, B.~W. (1982).
\newblock {Hydrogen-driving and the blue edge of compositionally stratified ZZ
  Ceti star models}.
\newblock \emph{\apjl} 252, L65--L68.
\newblock \doi{10.1086/183721}
\bibAnnoteFile{1982ApJ...252L..65W}

\bibitem[{{Wood}(1990)}]{1990PhDT.........5W}
{Wood}, M.~A. (1990).
\newblock \emph{{Astero-Archaeolgy Reading the Galactic History Recorded in the
  White Dwarf Stars}}.
\newblock Ph.D. thesis, Texas Univ., Austin.
\bibAnnoteFile{1990PhDT.........5W}

\bibitem[{{Woosley} and {Heger}(2015)}]{2015ApJ...810...34W}
{Woosley}, S.~E. and {Heger}, A. (2015).
\newblock {The Remarkable Deaths of 9-11 Solar Mass Stars}.
\newblock \emph{\apj} 810, 34.
\newblock \doi{10.1088/0004-637X/810/1/34}
\bibAnnoteFile{2015ApJ...810...34W}

\bibitem[{{Wu} and {Goldreich}(2001)}]{2001ApJ...546..469W}
{Wu}, Y. and {Goldreich}, P. (2001).
\newblock {Gravity Modes in ZZ Ceti Stars. IV. Amplitude Saturation by
  Parametric Instability}.
\newblock \emph{\apj} 546, 469--483.
\newblock \doi{10.1086/318234}
\bibAnnoteFile{2001ApJ...546..469W}

\bibitem[{{York} et~al.(2000){York}, {Adelman}, {Anderson}, {Anderson},
  {Annis}, {Bahcall} et~al.}]{2000AJ....120.1579Y}
{York}, D.~G., {Adelman}, J., {Anderson}, J.~E., Jr., {Anderson}, S.~F.,
  {Annis}, J., {Bahcall}, N.~A., et~al. (2000).
\newblock {The Sloan Digital Sky Survey: Technical Summary}.
\newblock \emph{\aj} 120, 1579--1587.
\newblock \doi{10.1086/301513}
\bibAnnoteFile{2000AJ....120.1579Y}

\bibitem[{{Zong} et~al.(2016{\natexlab{a}}){Zong}, {Charpinet}, and
  {Vauclair}}]{2016A&A...594A..46Z}
{Zong}, W., {Charpinet}, S., and {Vauclair}, G. (2016{\natexlab{a}}).
\newblock {Signatures of nonlinear mode interactions in the pulsating hot B
  subdwarf star KIC 10139564}.
\newblock \emph{\aap} 594, A46.
\newblock \doi{10.1051/0004-6361/201629132}
\bibAnnoteFile{2016A&A...594A..46Z}

\bibitem[{{Zong} et~al.(2016{\natexlab{b}}){Zong}, {Charpinet}, {Vauclair},
  {Giammichele}, and {Van Grootel}}]{2016A&A...585A..22Z}
{Zong}, W., {Charpinet}, S., {Vauclair}, G., {Giammichele}, N., and {Van
  Grootel}, V. (2016{\natexlab{b}}).
\newblock {Amplitude and frequency variations of oscillation modes in the
  pulsating DB white dwarf star KIC 08626021. The likely signature of nonlinear
  resonant mode coupling}.
\newblock \emph{\aap} 585, A22.
\newblock \doi{10.1051/0004-6361/201526300}
\bibAnnoteFile{2016A&A...585A..22Z}

\end{thebibliography}

\end{document}